\documentclass[11pt,aps,prd,onecolumn,nofootinbib,showkeys]{revtex4-2}
\usepackage{graphicx,epsfig,color}
\usepackage{bm}
\newcommand{\bc}{\begin{center}}
\newcommand{\ec}{\end{center}}
\newcommand{\be}{\begin{equation}}
\newcommand{\ee}{\end{equation}}
\newcommand{\bea}{\begin{eqnarray}}
\newcommand{\eea}{\end{eqnarray}}
\newcommand{\ba}{\begin{array}}
\newcommand{\ea}{\end{array}}
\newcommand{\lb}{\label}
\newcommand{\rf}{\ref}
\newcommand{\bfg}{\begin{figure}[htbp]}
\newcommand{\efg}{\end{figure}}

\begin{document}

\title{The interplay between compact and molecular structures
\protect \\ in tetraquarks}


\author{Hagop Sazdjian}
\email[]{sazdjian@ijclab.in2p3.fr}
%\thanks{}
\affiliation{Universit\'e Paris-Saclay, CNRS/IN2P3, IJCLab, 91405
Orsay, France}


%\date{\today}

\begin{abstract}
Due to the cluster reducibility of multiquark operators,
a strong interplay exists in tetraquarks between the compact
structure, resulting from the direct confining forces acting on
quarks and gluons, and the molecular structure, dominated by the
mesonic clusters. This issue is studied within an effective field
theory approach, where the compact tetraquark is treated as an
elementary particle. The key ingredient of the analysis is provided
by the primary coupling constant of the compact tetraquark to the
two mesonic clusters. Under the influence of this coupling, an
initially formed compact tetraquark bound state evolves 
towards a new structure, where a molecular configuration is also
present. In the strong-coupling limit, the evolution may end with
a shallow bound state of the molecular type. The strong-coupling
regime is also favored by the large-$N_c^{}$ properties of QCD.
The interplay between compact and molecular structures may provide
a natural explanation of the existence of many shallow bound states.
\end{abstract}


\maketitle

\section{Introduction and summary} \lb{s1}

The experimental discoveries, during the last two decades, of many
new particle candidates, corresponding to ``exotic hadrons''
\cite{Choi:2003ue,Aubert:2003fg,Besson:2003cp,Aubert:2005rm,
Ablikim:2013mio,Liu:2013dau,Ablikim:2013wzq,Aaij:2014jqa,Aaij:2015tga,
Aaij:2020fnh,Aaij:2020ypa,Wu:2020hmk}, 
not fulfilling the scheme of the standard quark model  
\cite{GellMann:1964nj,Zweig:1964jf,Jaffe:1976ig,Jaffe:2008zz}, has
given rise to thorough
theoretical investigations for the understanding of the nature and
structure of these states; recent review articles can be found in
\cite{Chen:2016qju,Hosaka:2016pey,Lebed:2016hpi,Esposito:2016noz,
Ali:2017jda,Guo:2017jvc,Olsen:2017bmm,Karliner:2017qhf,
Albuquerque:2018jkn,Liu:2019zoy,Ali:2019roi,Brambilla:2019esw,
Lucha:2021mwx}.
\par
The theoretical issue faced by exotic hadrons, also called
``multiquark states'', is whether they are formed, like ordinary
hadrons, by means of the confining forces that act on the quarks
and gluons, or whether they are formed, like molecular states, by
means of the effective forces that act on ordinary hadrons
\cite{Jaffe:2008zz}. (The term ``molecule'' refers here to the
color-neutral character of hadrons, in analogy with the molecules
formed by atoms \cite{Voloshin:1976ap,DeRujula:1976zlg}.)
In the former case, multiquark states are
expected to be compact objects, while in the latter case, they are
expected to be loosely bound states. For the formation of compact
multiquark states, the diquark model, in which two quarks
form a preliminary tight system, provides the simplest mechanism
to reach that goal \cite{Jaffe:2003sg,Shuryak:2003zi,Maiani:2004vq,
Maiani:2005pe}. Molecular-type states
\cite{Weinberg:1965zz,Voloshin:1976ap,DeRujula:1976zlg,
Tornqvist:1993ng}, also called ``hadronic molecules'', are studied
by means of effective field theories, based on approximate symmetry
properties and nonrelativistic approximation
\cite{Weinberg:1978kz,Gasser:1983yg,Manohar:1996cq,Georgi:1990um,
Neubert:1993mb,Casalbuoni:1996pg,Caswell:1985ui,Brambilla:2004jw}.  
\par
The reason these two competing alternatives are arising is related
to the fact that the multiquark operators that generate multiquark
states are not color-irreducible, in contrast to the ordinary hadron
case, in the sense that they are decomposable along combinations of
clusters of ordinary hadron operators
\cite{Jaffe:2008zz,Lucha:2019cdc}. There are, therefore, two different
ways of considering the construction of a multiquark state, as
depicted above. The main issue is which one reproduces the most
faithful description of reality. The existence or emergence of
hadronic clusters inside a multiquark state might be an indication
of a kind of instability when the state is built out of confining
forces. It might, at some stage, dislocate into the clusters, or,
in the case of a bound state, evolve towards another state,
dominated by the clusters, that is, towards a molecular-type state.
\par
A theoretical hint to analyze the problem is provided by the
study of the energy balance of the two types of configuration
\cite{Lucha:2021mwx}.
This is most easily done for heavy or static quarks, for which
lattice calculations are available \cite{Dosch:1982ep,
Alexandrou:2004ak,Okiharu:2004wy,
Okiharu:2004ve,Suganuma:2011ci,Cardoso:2011fq,Bicudo:2017usw}.
In the strong coupling limit of lattice theory, analytic expressions
are obtained by means of Wilson-loop expectation values, which
satisfy the area law, or, more generally, are saturated by minimal
surfaces; the corresponding predictions have been verified by
direct numerical calculations on the lattice (cf. previous
references).
The qualitative result that emerges from the latter calculations
is the following. When the quarks and antiquarks of the multiquark
state are gathered into a small volume, it is the compact multiquark
configuration that is energetically favored, while in situations
where the quarks and antiquarks are seperated from each other at
larger distances, it is the cluster-type configurations that are
energetically favored. Therefore, there are nonzero probabilities
for each type of configuration to occur for the description of the
multiquark state. However, since quarks and antiquarks are moving
objects and generally reaching, even with small probabilities, large
distances, whose integrated volume may be much larger than the
small volume of the compact configuration, one expects that an
initially formed compact state would gradually evolve towards a
cluster-type configuration, typical of a molecular state. In
coordinate space, the core of the multiquark state would be better
described by the compact representation, whereas the outer layer
would be better described by the molecular representation. The
relative weight of each representation would, of course, depend on
specific parameters, such as the quark masses and the quantum
numbers that are involved\footnote{
This scheme had been foreseen in the past by Manohar and Wise
\cite{Manohar:1992nd}, who have predicted, in the
presence of two heavy quarks and on the basis of the properties of
the confining interactions at short distances, the existence of a
tetraquark bound state. They, however, recognized that the
large-distance dynamics should be better described by meson-meson
interactions and switched for the description of that domain to
chiral perturbation theory.}.
The above results have led, in spectroscopic calculations, to the
introduction of the concept of ``configuration-space-partitioning''
(or for short ``geometric partitioning'') which is realized in the
so-called ``flip-flop'' potential model 
\cite{Miyazawa:1979vx,Lenz:1985jk,Oka:1984yx,Oka:1985vg,Carlson:1991zt,
Martens:2006ac,Vijande:2007ix,Richard:2009rp,Ay:2009zp,Vijande:2011im,
Bicudo:2010mv,Bicudo:2015bra,Bicudo:2015kna}, which takes into account
more faithfully the role played by each configuration in the
formation of multiquark states. However, because of the complicated
nature of the constraints, which are coordinate dependent, this model,
apart from simplified cases, has not yet led to full spectroscopic
results, to be compared, on quantitative grounds, with experimental
data.
\par
In principle, if the tetraquark bound state problem could have been
solved with high precision, taking into account all interactions
that act between quarks and gluons, one would obtain the exact
knowledge about its structure. Unfortunately, this is not currently
the case; one is obliged to adopt approximations and proceed step by
step by including additional inputs to improve the predictions.
As mentioned above, the simplest approximations are either the
compact scheme or the molecular scheme. While the latter scheme,
based on hadron-hadron interactions, has a sufficiently developed
theoretical background, the former one needs further analysis.
In the diquark model, the diquark being considered in particular
in its color-antisymmetric representation (ignoring here spin
degrees of freedom) within a very small volume (pointlike or almost
pointlike approximation), one always has tetraquark (or multiquark)
bound states \cite{Ebert:2007rn,Ebert:2010af,Faustov:2021hjs,
Ali:2017wsf,Ali:2018xfq,Lebed:2017min,Giron:2020qpb}; this is due
to the fact that, in that approximation, all forces acting on the
various small volumes (or points) are of the attractive confining  
types. This is not the case of the molecular scheme, where the
occurrence of a bound state depends on the strength of the
attractive forces. Therefore, in the compact
scheme, one is entitled to start with a tetraquark (or multiquark)
candidate, with all its accompanying multiplicities. The main
problem that is encountered here is the evaluation of the effect
the mesonic (or hadronic) clusters could have on that bound state.
The key ingredient which enters in the description of that effect
is the effective coupling constant of the compact tetraquark to
the meson clusters. One easily guesses that stronger the
latter quantity is, more important is the transformation of the
compact state into a cluster-like state, which ultimately might
take the appearance of a molecular-type state.
\par
It is the main objective of the present paper to evaluate the
interplay between the compact and molecular structures of possibly
existing tetraquark states. For this, we shall adopt methods of
effective field theories, remaining at the same time at the level
of simple qualitative features.
\par
In effective field theories of mesons, interactions are described
by meson exchanges and by contacts. At lower energies,
the exchanged meson fields can be integrated out and one remains only
with a theory with contact-type interactions
\cite{Weinberg:1991um,Luke:1996hj},
which we call here the ``lower-energy'' theory. The correspondence
between the parameters of the two types of theory is not, however,
simple and a physical understanding of the results necessitates a
more detailed investigation. We devote Sec.~\rf{s2} to a
presentation of this aspect of the problem. Taking into account
the various physical conditions and known results, we propose, in
Sec.~\rf{s3}, an empirical formula, which relates, in an explicit
way, the coupling constant of the lower-energy theory to that of the
Yukawa-type theory, and allows an easy understanding of the conditions
in which a bound state may emerge. In order to emphasize the
qualitative features of the approach, we neglect spin effects and
limit ourselves to scalar interactions with scalar particles,
ultimately considered in the nonrelativistic limit.
The resulting effective theory is then used
to study the meson-meson interaction through the scattering
amplitude and the determination of the possibly existing bound
state properties. The corresponding scattering length and effective
range are evaluated in Sec.~\rf{s4}. Some of the results of
Secs.~\rf{s3} and \rf{s4} are well known in the literature and
are presented here for the purpose of introducing the method of
approach that is applied for more general cases. 
\par
The case of compact tetraquarks is studied in Sec.~\rf{s5}.
With respect to the meson clusters, the compact tetraquark can
be represented, in first approximation, as an elementary particle,
whose internal structure would be relevant only at short-distance
scales. It is then essentially characterized by its mass and quantum
numbers and described by means of its propagator. The tetraquark,
because of its internal structure, has necessarily interactions with
meson pairs, and in particular with those lying closest to its mass.
In the simplest case of one meson-pair, one may introduce a bare
coupling constant for the interaction tetraquark-two-mesons and
analyze its influence on the properties of the tetraquark through
the radiative corrections it induces. For the bound state case, it
is assumed that the bare compact tetraquark mass lies below the
two-meson threshold. It turns out that, in general, the
compact-tetraquark--two-meson interaction shifts the binding
energy of the tetraquark to lower values. In the strong coupling
limit, the shift may even transform
the compact tetraquark into a shallow bound state, typical of loosely
bound hadronic molecules. This phenomenon is best represented by
means of the ``elementariness'' parameter $Z$, introduced by
Weinberg \cite{Weinberg:1965zz}, which measures the probability
of a bound state to be considered as elementary, the complementary
quantity, $(1-Z)$, representing its ``compositeness''. In the
strong coupling limit, described above, $Z$ takes small values,
approaching zero. In parallel, the physical coupling constant
of the tetraquark to two mesons tends also to zero in the same
limit. The value of $Z$ is also measured by means of the scattering
length and the effective range parameter, the latter taking negative
values when $Z\neq 0$.
\par
These results, which are the main outcome of the present paper,
provide a more refined understanding of the structure of observed
tetraquark candidates. Many shallow bound states, which are
typical of molecular states, might have a compact origin, provided
$Z\neq 0$. They would be the result of the deformation, under the
influence of the mesonic clusters, of the initially formed compact
tetraquark. This is another illustration of the dual
representation of the tetraquark found in lattice theory on the
basis of the energy balance analysis
\cite{Dosch:1982ep,Alexandrou:2004ak,Okiharu:2004wy,
Okiharu:2004ve,Suganuma:2011ci,Cardoso:2011fq,Bicudo:2017usw}.
\par
On more general grounds, shallow bound states are usually
considered as belonging to universality classes, whose binding
energy values are not naturally explained by means of the
interaction scales of the system \cite{Braaten:2004rn}.
The previous results bring a new lighting to that problem in the
case of tetraquarks. The coupling constant
compact-tetraquark--meson-clusters introduces an additional scale
parameter, whose strong-coupling limit naturally explains the origin
of the shallowness.
\par
The case of resonances occurs when the bare mass of the compact
tetraquark lies above the two-meson threshold. Here, however,
contrary to the bound state case, additional constraints appear
for the existence of a physical resonance. The latter may exist
only in the weak-coupling regime. Large values of the coupling
constant resend the state to the bound state domain, while for
a finite interval of the coupling constant, occurring prior to the
strong-coupling regime, the tetraquark state may disappear from
the spectrum.
\par
Section \rf{s6} brings complementary informations with respect to
Sec.~\rf{s5}, by also considering, in addition to the
tetraquark-two-meson interaction effect, the influence of
meson-meson interactions, which now renormalize the primary (bare)
coupling constant and introduce a competing effect coming from
direct molecular-type forces. The qualitative conclusions drawn
in Sec.~\rf{s5} remain, however, valid.
Section \rf{s7} is devoted to an analysis of the problem in the
large-$N_c^{}$ limit of QCD. In that limit, the theory provides
additional support to the dominance of the strong-coupling regime
in the effective interaction compact-tetraquark--meson-clusters. 
Conclusion follows in Sec.~\rf{s8}. A few detailed analytic
expressions, approximating energy eigenvalues, are gathered in
the Appendix.
\par

\section{Reduction to contact-type interactions} \lb{s2}

Effective field theories, which result from the integration of fields
operating mainly at high energies, are generally characterized by the
presence of contact-type interactions. Often, depending on the energy
scale that is considered, these coexist with ordinary-type
interactions, whose prototype is the Yukawa interaction, responsible
of meson exchanges between interacting particles. At lower energies,
the exchange-meson fields themselves are integrated out and one remains
only with contact-type interactions. A representative example of such
a theory is chiral perturbation theory
\cite{Weinberg:1978kz,Gasser:1983yg,Manohar:1996cq}, in which the
interacting particles are the pseudo-Goldstone bosons and where
all other massive particle fields have been integrated out.
However, explorations of more refined properties, related to
spectroscopic problems and to the physical interpretation of numerical
values of parameters, may require a more detailed knowledge of the
connection between the lower-energy theory and its higher-energy
generator.
\par

\subsection{Spectroscopic properties of the higher-energy theory}
\lb{s21}

To illustrate the above aspect of the question, we shall consider,
following Reference \cite{Luke:1996hj}, a higher-energy theory,
where two massive scalar particles, with masses $m_1^{}$ and $m_2^{}$,
interact by means of the exchange of a scalar particle with mass
$\mu$, described by the interaction Lagrangian density
\be \lb{2e1}
\mathcal{L}_I^{}=\sum_{i=1}^22m_ig\phi_i^{\dagger}\phi_i^{}\varphi,
\ee
where, for simplicity, after factorizing, for dimensionality reasons,
the mass terms $2m_i^{}$, we have chosen the same (dimensionless)
coupling constant $g$ for the two particles. The fields $\phi_i$
($i=1,2$) correspond to those of the external massive particles,
while $\varphi$ corresponds to the exchanged particle field.
\par
The scattering amplitude of the process
$\{1(p_1')+2(p_2')\rightarrow 1(p_1^{})+2(p_2^{})\}$ is given,
in lowest order in $g$, by the Born term, or the ladder diagram:
\be \lb{2e2}
i\mathcal{T}=-i\frac{4m_1m_2g^2}{q^2-\mu^2+i\epsilon},
\ee
where $q$ is the momentum transfer: $q=p_1^{}-p_1'=p_2'-p_2^{}$.  
This term iteratively generates higher-order diagrams, representing
the series of ladder-type diagrams, which plays a basic role in a
possible production of bound states. They are represented in
Fig.~\rf{f1}.
\bfg
%\vspace*{1 cm}
\bc
\epsfig{file=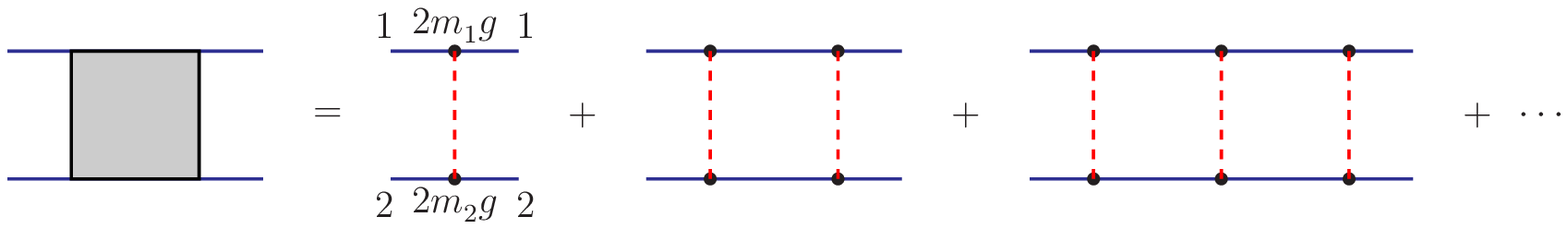,scale=0.7}
\caption{The meson-meson scattering amplitude in terms of the series
of ladder diagrams.}
\lb{f1}
\ec
\efg
\par
A complete study of the theory requires the evaluation of the effects
of self-energy and vertex corrections. These, however, do not play a
fundamental role in the extraction of the main qualitative features
that are of interest for our purpose and will be ignored.
\par
The bound state problem in the nonrelativistic limit can be studied
by means of the Schr{\"o}dinger equation.
The latter is obtained from the Bethe--Salpeter equation, with the
kernel considered in the ladder approximation, represented by the
right-hand side of Eq.~(\rf{2e2}). One has first to use the
instantaneous approximation, where one neglects, in the cm frame,
the temporal component of the momentum transfer $q$, and then to take
the nonrelativistic limit. The Schr{\"o}dinger equation in $x$-space
is:
\be \lb{2e3}
E\psi(\mathbf{x})=\Big(\frac{\mathbf{p}^2}{2m_r^{}}
+V(r)\Big)\psi(\mathbf{x}),\ \ \ r=\sqrt{\mathbf{x}^2}, \ \ \ 
\mathbf{x}=\mathbf{x}_1^{}-\mathbf{x}_2^{},\ \ \
m_r^{}=\frac{m_1^{}m_2^{}}{(m_1^{}+m_2^{})},
\ee
where $V(r)$ is the well-known Yukawa potential:
\be \lb{2e4}
V(r)=-(\frac{g^2}{4\pi})\frac{1}{r}e^{{\displaystyle -\mu r}}.
\ee
Making in (\rf{2e3}) the changes of variable and parameter
\cite{Carbonell:2012bv}
\be \lb{2e5}
\mathbf{x}=\mathbf{x}'/\mu,\ \ \ \ \mathbf{p}=\mu\mathbf{p'},
\ \ \ \ E=\frac{\mu^2}{2m_r^{}}E',\ \ \ \
g^2=\overline{g}^2\frac{\mu}{2m_r^{}},
\ee
one reduces the Schr{\"o}dinger equation to a dimensionless equation
with a single parameter, $\overline{g}^2$:
\be \lb{2e6}
E'\psi(\mathbf{x}')=\Big(\mathbf{p}^{\prime 2}
-(\frac{\overline{g}^2}{4\pi})\frac{1}{r'}e^{{\displaystyle -r'}}
\Big)\psi(\mathbf{x}').
\ee
\par
Compared to the Coulomb potential, the Yukawa potential (\rf{2e4})
is of the short-range type and hence not all values of
$\overline{g}^2$ may produce bound states. There exists a critical
value of it, $\overline{g}_{\mathrm{cr}}^2=1.68\times (4\pi)$
\cite{Luke:1996hj,Carbonell:2012bv}, below which bound states do not
exist. Their existence is ensured by the inequality
\be \lb{2e7}
(\frac{\overline{g}^2}{4\pi})\ge (\frac{\overline{g}_{\mathrm{cr}}^2}
{4\pi})=1.68.
\ee
At the critical value of the coupling constant, a zero-energy bound
state appears. When $\overline{g}^2$ increases, gradually new bound
states appear, while the ground state binding energy itself
increases. 
\par
In more general cases, one may have the sum of several Yukawa
potentials with different meson exchanges. In such cases, one loses
the notion of a universal coupling constant squared $\overline{g}^2$.
To continue exploring qualitative aspects of the problem, it would
be advantageous to approximate the sum by a single Yukawa potential
with a mean exchanged mass and a mean coupling constant.
\par
General qualitative properties of the Yukawa potential, concerning
the related spectroscopy and the poles of the corresponding $S$-matrix,
could be obtained by considering the soluble model of the spherically
symmectric rectangular potential well, which is a prototype of
short-range potentials \cite{Nussenzveig:1959,vanKolck:1998bw}.
If $V_0^{}$ is the depth of the potential and $R$ its width, then
there is a correspondence between $\overline{g}^2/(4\pi)$ and the
product $2m_r^{}V_0^{}R^2\equiv A^2$. The critical value of $A^2$ is
equal to $\pi^2/4=2.46$, which is of the same order of magnitude as
$\overline{g}_{\mathrm{cr}}^2/(4\pi)$ [Eq.~(\rf{2e7})].
\par

\subsection{The lower-energy effective field theory} \lb{s22}

A lower-energy effective field theory can be obtained by integrating
out the mediator field $\varphi$. One then obtains a theory where the
only remaining fields are the $\phi_i$s. Their interactions are
represented by an infinite series of contact terms with increasing
powers of products of $\phi_i$s, also possibly containing derivative
couplings. They are classified according to their dimensionality and
a systematic expansion is organized according to definite power
counting rules. The coupling constants of the lower-energy theory
are usually determined from matching conditions of the scattering
amplitudes calculated in the two theories. At low energies, it is
the lowest-dimension operator that is expected to provide the leading
contribution. We stick in the following to that term. The
corresponding Lagrangian density, for the mutual interaction of
particles 1 and 2, is 
\be \lb{2e8}
\mathcal{L}_{I,\mathrm{eff}}^{}=h\phi_1^{\dagger}\phi_1^{}
\phi_2^{\dagger}\phi_2^{},
\ee
where $h$ is the coupling constant. This term, like (\rf{2e1}),
generates by iteration a chain of loop or bubble diagrams, which
are represented in Fig.~\rf{f2}.
\bfg
%\vspace*{1 cm}
\bc
\epsfig{file=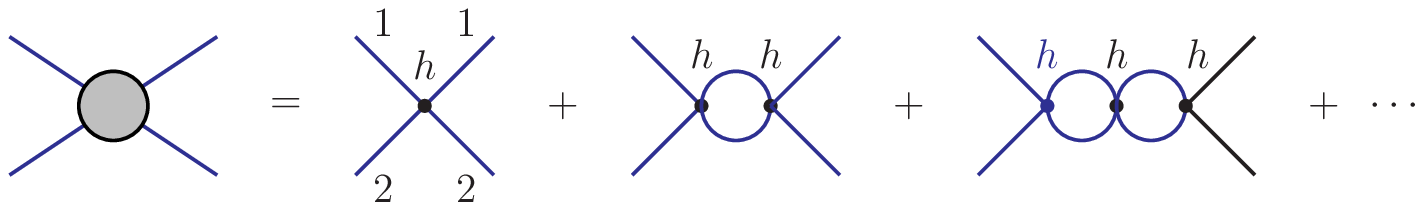,scale=0.8}
\caption{The meson-meson scattering amplitude in terms of the chain
of loop diagrams generated by the contact term (\rf{2e8}).}
\lb{f2}
\ec
\efg
\par
The problem that interests us is to what extent the lower-energy
theory in its leading-order approximation, represented by the
Lagrangian density (\rf{2e8}), provides a faith\-ful de\-scrip\-tion of
the higher-energy theory, considered in the ladder-app\-roxi\-ma\-tion,
in the bound state domain.
\par
Similar searches as above have been undertaken long ago in the past,
although with a different viewpoint. In this respect, Reference
\cite{Lurie:1964}, provides a rather detailed account of the
corresponding approach. The objective has been to establish an
equivalence theorem between the two theories described by Lagrangian
densities of the types of (\rf{2e1}) and (\rf{2e8}), respectively,
the external particles being here fermions. Putting aside the
question of the non\-renor\-mali\-zabi\-lity of the four-Fermi
interaction theory, the approach has consisted in considering the
first ladder diagram of Fig. \rf{f1}, together with the radiative
corrections of the exchanged meson propagator, and searching for
conditions of equivalence of the chain of diagrams of Fig. \rf{f2},
considered in the $t$-channel. Our line of investigation is rather
different, being based on the effective field theory approach,
searching for matching conditions for the leading terms of the
scattering amplitude in the $s$-channel. 
\par
Coming back to the present approach with its matching conditions,
it turns out that all diagrams of Fig. \rf{f1}, besides contributing
to the coupling constants of higher-dimensional operators, also give
contributions to the coupling constant $h$ of (\rf{2e8}). Therefore,
$h$, considered as a function of $g^2$ (or $\overline{g}^2$), has an
expression in the form of an infinite series of it:
\be \lb{2e9}
h=\sum_{n=0}^{\infty}h^{(n)},\ \ \ \ \ \ \ \ \ 
h^{(0)}=\frac{4m_1^{}m_2^{}}{\mu^2}g^2,
\ee
where $h^{(n)}$ represents the contribution coming from the $n$-loop
diagram (proportional to
$(\frac{\overline{g}^2}{4\pi})^{n}h^{(0)}$).
Notice that $h^{(0)}$ is positive.
While the first few terms of the series of $h$ are explicitly
calculable \cite{Luke:1996hj}, the complete knowledge of the
expression of $h$ in terms of $g^2$ is not known. It is evident, from
the lower-bound (\rf{2e7}), that, for the bound state problem, one
cannot be satisfied by a perturbative expansion of $h$, but rather
a full nonperturbative expression of it will be needed in order
to understand and interpret the bound state properties of the
lower-energy theory in terms of those of the higher-energy theory.
\par

\section{Nonperturbative properties of the coupling constant of
the low-energy theory} \lb{s3}

The nonperturbative properties of the coupling constant $h$ can be
deduced from the study of the bound state problem of the low-energy
theory. The result of the latter problem is well known in the 
literature \cite{Weinberg:1991um,Luke:1996hj} and we briefly
sketch the corresponding procedure. The scattering amplitude
resulting from the chain of diagrams of Fig. \rf{f2} is a
geometric series expressible in terms of the two-point loop
integral $J(s)$:
\bea
\lb{3e1}
& &\mathcal{T}\equiv\mathcal{T}_{\mathrm{eff}}^{}
=\frac{h}{1-ihJ(s)},\ \ \ \ \ \ s=P^2=(p_1^{}+p_2^{})^2,\\
\lb{3e2}
& &J(s)=\int\frac{d^4k}{(2\pi)^4}
\frac{i}{(p_1^{}+k)^2-m_1^2+i\epsilon}
\frac{i}{(p_2^{}-k)^2-m_2^2+i\epsilon}.
\eea
$J(s)$ is divergent, but its divergence can be absorbed in $h$ by
renormalization. Defining $J^{\mathrm{r}}$ as the regularized
(finite) part of $J$ and $J^{\mathrm{div}}$ as its diverging part,
such that
\be \lb{3e3}
J(s)=J^{\mathrm{div}}+J^{\mathrm{r}}(s),
\ee
one obtains
\be \lb{3e4}
\frac{1}{h}-iJ^{\mathrm{div}}=\frac{1}{h^{\mathrm{r}}}
\ \ \ \ \ \ \ \Longleftrightarrow \ \ \ \ \ \ \
\frac{h}{1-ihJ(s)}=
\frac{h^{\mathrm{r}}}{1-ih^{\mathrm{r}}J^{\mathrm{r}}(s)},
\ee
where $h^{\mathrm{r}}$ is the renormalized finite part of $h$.
Notice that this renormalization implies that the unrenormalized $h$
is a vanishing quantity:
\be \lb{3e4a}
h=\frac{h^{\mathrm{r}}}{1+ih^{\mathrm{r}}J^{\mathrm{div}}}.
\ee
In the limit where, for finite $h^{\mathrm{r}}$,
$iJ^{\mathrm{div}}\rightarrow \infty$, one has $h\rightarrow 0$.
\par
Regularizing $J$ by dimensional regularization and keeping in
the diverging part only mass-independent terms, $J^{\mathrm{r}}$
takes the following form:
\be \lb{3e5}
J^{\mathrm{r}}(s)=\frac{i}{16\pi^2}\,\Big[\,
\ln\Big(\frac{m_1^{}m_2^{}}{4\pi\overline\mu^2}\Big)
+\frac{(m_1^2-m_2^2)}{2s}\ln\Big(\frac{m_1^2}{m_2^2}\Big)+Q(s)\,\Big].
\ee
$\overline \mu$ is the mass scale introduced by the
dimensional regularization, whose value can be chosen at will;
$Q(s)$ is defined in the complex $s$-plane, cut on the real
axis from $s=(m_1^{}+m_2^{})^2$ to $+\infty$ and from
$s=(m_1^{}-m_2^{})^2$ to $-\infty$, and has the expression
\be \lb{3e6}
Q(s)=\frac{\sqrt{\lambda(s)}}{s}\,\ln\Big
(\frac{\sqrt{s-(m_1^{}+m_2^{})^2}
+\sqrt{s-(m_1^{}-m_2^{})^2}}{\sqrt{s-(m_1^{}+m_2^{})^2}
-\sqrt{s-(m_1^{}-m_2^{})^2}}\Big),
\ee  
where
\be \lb{3e7}
\lambda(s)\equiv\lambda(s,m_1^2,m_2^2)=\Big(s-(m_1^{}+m_2^{})^2\Big)
\Big(s-(m_1^{}-m_2^{})^2\Big).
\ee
(The square-roots are defined for complex $s$ as
$\sqrt{s-m^2}=e^{i\alpha/2}\sqrt{|s-m^2|}$, where $\alpha$ is the angle
made by $(s-m^2)$ with the positive real axis starting from $m^2$.)
On the real axis, $Q(s)$ takes the following forms: 
\be \lb{3e6a}
Q(s)=\left\{
\ba{l}
-\frac{\sqrt{\lambda(s)}}{s}\ln\Big(\frac{\sqrt{(m_1^{}+m_2^{})^2-s}
+\sqrt{(m_1^{}-m_2^{})^2-s}}{\sqrt{(m_1^{}+m_2^{})^2-s}
-\sqrt{(m_1^{}-m_2^{})^2-s}}\Big),\ \ \ \
s<(m_1^{}-m_2^{})^2,\\
+\frac{\sqrt{-\lambda(s)}}{s}\Big
[\pi-2\arctan\Big(\frac{\sqrt{(m_1^{}+m_2^{})2-s}}
{\sqrt{s-(m_1^2-m_2^2)}}\Big)\Big],\ \ \ \
(m_1^{}-m_2^{})^2<s<(m_1^{}+m_2^{})^2,\\
+\frac{\sqrt{\lambda(s)}}{s}\Big[\ln\Big
(\frac{\sqrt{s-(m_1^{}-m_2^{})^2}
+\sqrt{s-(m_1^{}+m_2^{})^2}}{\sqrt{s-(m_1^{}-m_2^{})^2}
-\sqrt{s-(m_1^{}+m_2^{})^2}}\Big)
\mp i\pi\Big],\\
\hspace{4. cm} Re(s)>(m_1^{}+m_2^{})^2,\ \ \ \ Im(s)=\pm\epsilon,
\ \ \epsilon>0.
\ea
\right.
\ee
(The square-roots in (\rf{3e6a}) are defined with positive values,
their arguments representing moduli. $J^{\mathrm{r}}(s)$ is finite
at $s=0$.)
Physical quantities should be independent of $\overline \mu$.
Since $\mathcal{T}_{\mathrm{eff}}^{}$ is such a quantity, this
implies that $h^{\mathrm{r}}$ itself should be $\overline \mu$
dependent and should cancel the $\overline \mu$-dependence of
$J^{\mathrm{r}}$. The absorption of the term
$\ln\Big(\frac{m_1^{}m_2^{}}{4\pi\overline\mu^2}\Big)$ into a
redefined $1/h^{\mathrm{r}}$, which amounts to also absorbing the
same quantity into $J^{\mathrm{div}}$, is the simplest way of
ensuring this.
We shall adopt henceforth that procedure.
It is advantageous to define $h^{\mathrm{r}}$ as having a
simple physical interpretation. A natural choice is the value
of $\mathcal{T}_{\mathrm{eff}}^{}$ at the two-particle threshold 
\cite{Weinberg:1991um}. We shall continue using the same notation
for the redefined $h^{\mathrm{r}}$ and $J^{\mathrm{r}}$; then
the redefined $J^{\mathrm{r}}$ obtains the
form
\be \lb{3e8}
J^{\mathrm{r}}(s)=\frac{i}{16\pi^2}
\Big[\,\Big(\frac{1}{s}-\frac{1}{(m_1^{}+m_2^{})^2}\Big)
\frac{(m_1^2-m_2^2)}{2}\ln\Big(\frac{m_1^2}{m_2^2}\Big)
+Q(s)\,\Big].
\ee
\par
The scattering amplitude (\rf{3e1}) then takes the form, according
to (\rf{3e4}),
\be
\lb{3e1a}
\mathcal{T}=\frac{h^{\mathrm{r}}}
{1-ih^{\mathrm{r}}J^{\mathrm{r}}(s)}.
\ee
\par
Bound states of $\mathcal{T}$ will be
identified as tetraquark states of the molecular type, whose
parameters and ingredients will be labeled by the indices $tm$;
they correspond to the solutions of the equation
\be \lb{3e9}
\frac{1}{h^{\mathrm{r}}}-iJ^{\mathrm{r}}(s_{tm}^{})=0.
\ee
(We shall often omit, for the simplicity of notation and when no
ambiguity is present, the index $\mathrm{r}$ from $h^{\mathrm{r}}$
and $J^{\mathrm{r}}$.)
We restrict ourselves to the case of possible nonrelativistic
solutions, located near the two-particle threshold. We introduce the
nonrelativistic energy $E_{tm}^{}$ through the definition
\be \lb{3e10}
\sqrt{s_{tm}^{}}=(m_1^{}+m_2^{})+E_{tm}^{},
\ee
and, to simplify notations, we shall use henceforth the reduced
dimensionless energy variable $e$ and bound state energy
$e_{tm}^{}$, respectively:
\be \lb{3e10a}
e\equiv\frac{E}{2m_r^{}},\ \ \ \ \ \ \ \
e_{tm}^{}\equiv\frac{E_{tm}^{}}{2m_r^{}},
\ee
where $m_r^{}$ is the reduced mass [Eq.~(\rf{2e3})].
Retaining, in the second of Eqs. (\rf{3e6a}), the leading
term in $\sqrt{-e_{tm}^{}}$, which is contained in its first piece, one
finds a unique possible solution:
\be \lb{3e11}
\sqrt{-e_{tm}^{}}=-\frac{16\pi}{\alpha h},\ \ \ \ \ \ \
\alpha\equiv\frac{4m_r^{}}{(m_1^{}+m_2^{})}.
\ee
The existence of the solution is conditioned by a negative value
of $h$. This means that $h$ must have changed sign with respect to
the perturbative expression $h^{(0)}$ [Eq.~(\rf{2e9})]. 
To ensure the nonrelativistic interpretation of the solution, one
must have a large value of $|h|$, such that
\be \lb{3e11a}
16\pi/|h|\ll 1.
\ee
When $|h|\rightarrow \infty$, the bound state approaches the
two-particle threshold. On the other hand, when $h\rightarrow 0$
with negative values, the binding energy increases and tends to
$\infty$.
\par
The stability of the solution (\rf{3e11}) in the presence of
higher-order terms in $\sqrt{-e_{tm}^{}}$ can be studied
by also considering in $J^{\mathrm{r}}(s)$ terms of order
$(-e)$, which are contained in the second term of
$Q(s)$ [Eq.~(\rf{3e6a})] and the remaining terms of
$J^{\mathrm{r}}(s)$ [Eq. (\rf{3e7})]. Accepting condition
(\rf{3e11a}), the resulting equation continues reproducing
the solution (\rf{3e11}), but yields also a second solution,
which lies far from the two-particle threshold and does not have
the required nonrelativistic limit. Solution (\rf{3e11}) remains
therefore the only stable nonrelativistic solution of
Eq.~(\rf{3e9}).
\par
The behavior of the bound state energy with respect to the
variations and the order of magnitude of the coupling constant
$h$ is unusual. Generally, if a coupling constant
of a theory increases up to infinity, one expects to find
instabilities or phase transitions, while here one finds a smooth
behavior in the vicinity of the two-particle threshold. Similarly,
vanishing values of the coupling constant should lead the theory
towards its perturbative regime. This means that the coupling
constant $h$ of the low-energy theory cannot be interpreted as an
elementary or an ordinary coupling constant. To interpret correctly
its role, one should try to connect it more explicitly to the original
coupling constant $g$ of the higher-energy theory. In the latter
theory, the lowest bound state approaches the two-particle threshold
when $g$ approaches its critical value $g_{\mathrm{cr}}^{}$ from above.
Therefore, negative values of $h$ correspond to values of $g$ greater
than $g_{\mathrm{cr}}^{}$. Large and negative values of $h$ would
correspond to the approach $g\rightarrow g_{\mathrm{cr}}^{}$, at which
value $h$ would have a singularity. Vanishing of $h$ with negative
values would correspond to the increasing of $g$ up to infinity.
When $g$ is lower than $g_{\mathrm{cr}}^{}$, $h$ should change sign and
the system would enter in a phase characterized by the absence of
bound states. Finally, the vanishing of $h$ with positive values would
correspond to the approach to the perturbative regime. A schematic
representation of these correspondences is shown in Fig.~\rf{f3}
\cite{Lucha:2021mwx}.
\bfg
%\vspace*{1 cm}
\bc
\epsfig{file=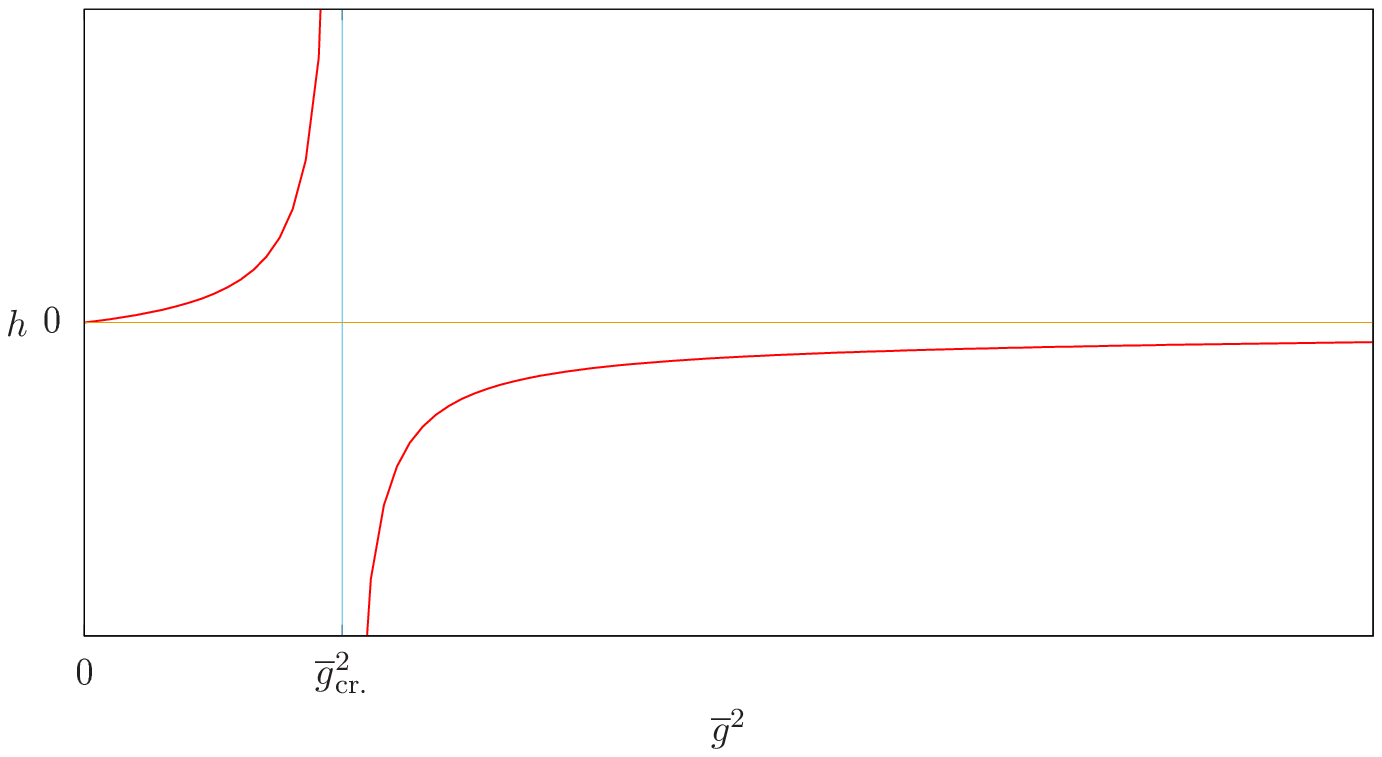,scale=0.8}
\caption{A schematic representation of the relationship between
the low-energy coupling constant $h$ and the high-energy coupling
constant $g$, the latter represented through $\overline{g}^2$
[Eq.~(\rf{2e5})].}
\lb{f3}
\ec
\efg
\par
The exact relationship between $h$ and $g$ not being available at
present, it would be useful to have an approximate or an empirical
relation which qualitatively reproduces its main properties and
allows for an easy understanding of the physical situation that
is considered. For this, we propose the following formula
between $h$ and $\overline{g}^2$ [Eq.~(\rf{2e5})], the latter
having a more universal meaning than $g^2$:
\be \lb{3e12}
\frac{\alpha h}{16\pi}=\frac{2m_r^{}}{\mu}
\frac{\overline{g}^2}{4\pi}
\frac{1}{(1-\overline{g}^2/\overline{g}_{\mathrm{cr}}^2)
(1+b\overline{g}^2/\overline{g}_{\mathrm{cr}}^2)},\ \ \ \ \ \
b\simeq 0.5,
\ee
where $b$ is an empirical parameter, whose approximate value has
been determined by numerical tests.
\par
This formula is expected to be approximately valid in the
nonrelativistic domain of the bound states. To test its validity,
we have compared the bound state energies, calculated from the
Schr{\"o}dinger equation [(\rf{2e3})-(\rf{2e6})] and from the
lower-energy theory [(\rf{3e11}) and (\rf{3e12})], in units of
$\mu^2/(2m_r^{})$. The results are graphically represented in
Fig.~\rf{f4}.
\bfg
%\vspace*{1 cm}
\bc
\epsfig{file=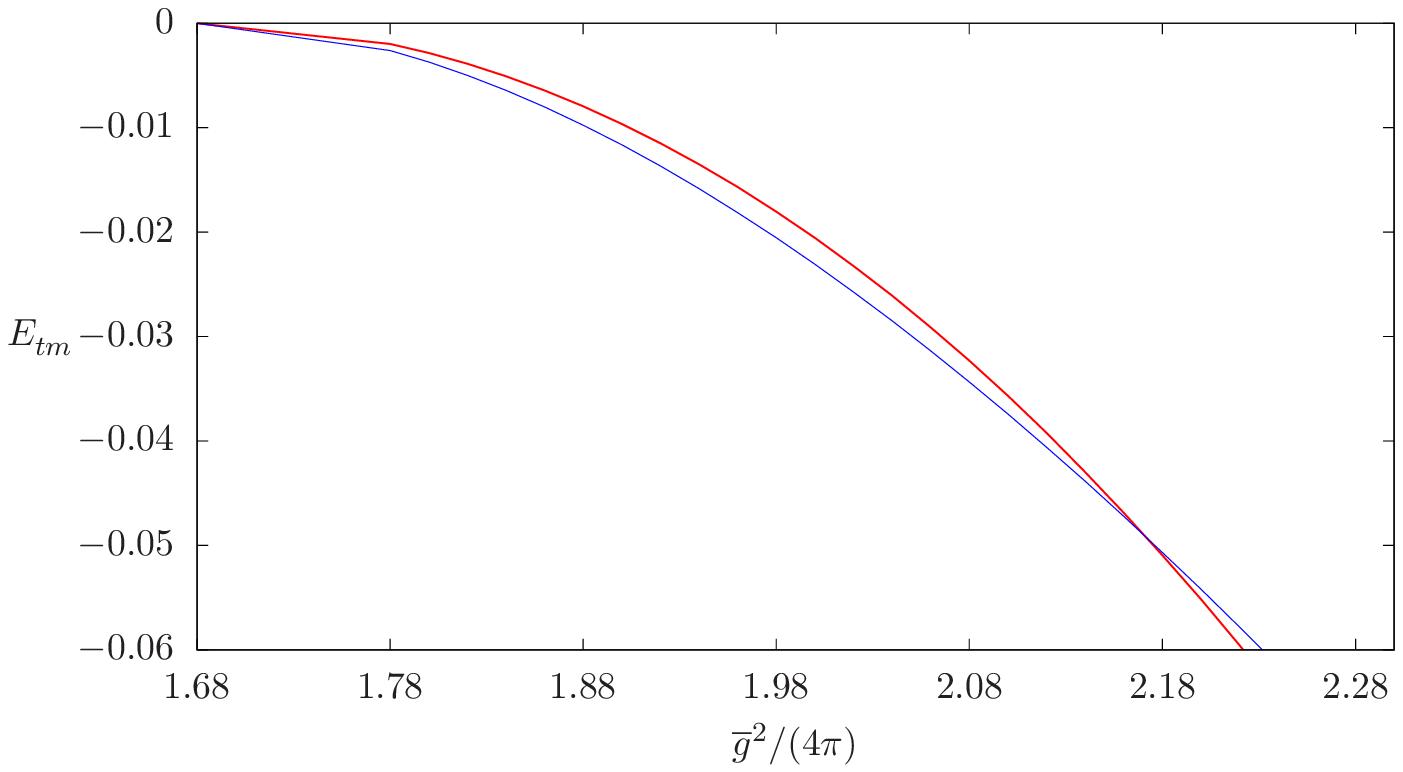,scale=0.8}
\caption{The bound-state energies calculated from the Yukawa
potential (red line) and from the lower-energy theory (blue 
line), in units of $\mu^2/(2m_r^{})$.}
\lb{f4}
\ec
\efg
One finds a rather satisfactory matching of the two predictions.
To have an idea of the values of the predicted binding energies,
choosing $2m_r^{}\simeq 2$ GeV (corresponding for instance to a
system of $D\overline D$ or $D^*\overline D$ mesons) and
$\mu\simeq 0.5-1.$ GeV (corresponding to the exchange of an
effective scalar meson), one has for the unit of energy
$\mu^2/(2m_r^{})\simeq 25-100$ MeV; the value 0.06 of $-E_{tm}^{}$ in
Fig.~\rf{f4} would correspond to a binding energy of $1.5-6$ MeV.
\par
By expanding in (\rf{3e12}) $\overline{g}^2$ around
$\overline{g}_{\mathrm{cr}}^2$, one obtains from (\rf{3e11}) the
further approximation of $e_{tm}^{}$ near the two-meson threshold:
\be \lb{3e13}
e_{tm}^{}\simeq -\Big(\frac{\mu}{2m_r^{}}\Big)^2\Big[
\Big(\frac{4\pi}{\overline{g}_{\mathrm{cr}}^2}\Big)(1+b)\Big]^2
\Big(\frac{\overline{g}^2}{\overline{g}_{\mathrm{cr}}^2}-1\Big)^2.
\ee
The quadratic behavior of $e_{tm}^{}$ with respect to the departure of
$\overline{g}^2$ from $\overline{g}_{\mathrm{cr}}^2$ is typical
of short-range potentials and can be verified on soluble models.
\par
As a side remark, let us notice that the extension of the formula
(\rf{3e12}) to the relativistic domain necessitates more elaborate
comparisons. The reason is that
in the latter domain one has no longer a single energy unit
and the corresponding generalization is not straightforward.
Furthermore, in the higher-energy theory, the Bethe--Salpeter
equation itself has difficulties to describe correctly the first
relativistic corrections in the ladder approximation with covariant
propagators \cite{Love:1977hj}; one is obliged to use either the
instantaneous approximation, or other quasipotential-type approaches,
which reduce the Bethe--Salpeter equation to a three-dimensional
equation. We shall be content, in the present work, to stick to
the nonrelativistic domain, where many experimental data still require
a detailed investigation.
\par
One can also obtain, from the expression (\rf{3e1a}) of the
scattering amplitude, the coupling constant of the bound state
to the constituent mesons, appearing in the residue of the bound
state pole. Designating by $M_1^{}$ and $M_2^{}$ the two constituent
mesons, the scattering amplitude has the following behavior near the
pole position of the bound state:
\be \lb{3e14}
\mathcal{T}\simeq -\frac{(m_1^{}+m_2^{})^2g_{TM_1^{}M_2^{}}^2}
{s-s_{tm}^{}},
\ee
where $g_{TM_1^{}M_2^{}}^{}$ represents the dimensionless coupling
constant of the tetraquark to the two mesons, defined with the
accompanying mass factor $(m_1^{}+m_2^{})$.
Expanding in (\rf{3e1a}) $J^{\mathrm{r}}(s)$ around $s_{tm}^{}$ and
using (\rf{3e9}), one obtains
\be \lb{3e15}
(m_1^{}+m_2^{})^2g_{TM_1^{}M_2^{}}^2=\frac{1}{iJ'(s_{tm}^{})},
\ee
giving
\be \lb{3e16}
g_{TM_1^{}M_2^{}}^2=32\pi\sqrt{-e_{tm}^{}}.
\ee
One notices that the coupling constant decreases when the bound
state approaches the two-particle threshold.
\par
Let us finally emphasize, as is evident from the previous results
and, in particular, from the structure of $\mathcal{T}$
[Eq.~(\rf{3e1})], that the low-energy theory can reproduce, in the
present scalar theory, only the ground state of the higher-energy
theory.
\par

\section{Scattering length and effective range} \lb{s4}

The expression of the scattering amplitude (\rf{3e1}) can also be
used in the scattering domain, where only the $S$-wave contributes.
Introducing the cm momentum
$k=\sqrt{\lambda(s,m_1^2,m_2^2)}/(2\sqrt{s})$
[Eq.~(\rf{3e7})] and expanding $J^{\mathrm{r}}(s)$ 
[(\rf{3e8}) and the third of Eqs.~(\rf{3e6a})], considered above
the cut, up to terms of order $k^2$, one finds
\be \lb{4e1}
\mathcal{T}=\frac{8\pi (m_1^{}+m_2^{})}
{\frac{8\pi (m_1^{}+m_2^{})}{h}+\frac{k^2}{\pi m_r^{}}(1-d)-ik},
\ee        
where we have defined
\be \lb{4e5}
d=\frac{1}{2}\Big(\frac{m_1^{}-m_2^{}}{m_1^{}+m_2^{}}\Big)
\ln(\frac{m_1^{}}{m_2^{}}).
\ee
On the other hand, $\mathcal{T}$ can be expressed in terms of the
$S$-wave phase shift $\delta_0^{}$ as
\be \lb{4e2}
\mathcal{T}=\frac{8\pi\sqrt{s}}{k\cot\delta_0^{}(k)-ik}.
\ee
The factor $\cot\delta_0^{}(k)$ is itself expressed through a
low-energy expansion in terms of the scattering length $a$ and the
effective range $r_e^{}$ \cite{Bethe:1949yr}:
\be \lb{4e3}
k\cot\delta_0^{}(k)=-\frac{1}{a}+\frac{1}{2}r_e^{}k^2,
\ee
yielding the identifications
\be \lb{4e4}
a=-\frac{h}{8\pi (m_1^{}+m_2^{})}=\frac{1}{2m_r^{}\sqrt{-e_{tm}^{}}},
\ \ \ \ \ \ \ r_e^{}=\frac{2}{\pi m_r^{}}(1-d).
\ee
(The relativistic correction coming from the expansion of $\sqrt{s}$
in the numerator of $\mathcal{T}$, Eq.~(\rf{4e2}), has been
neglected.) One finds that $a$ is proportional to
$-h$ and, therefore, has the same type of behavior as $-h$ in terms
of the coupling constant squared $\overline g^2$ (Fig.~\rf{f3}).
(This has also been shown in \cite{Carbonell:2012bv}.) According to
whether $\overline g^2$ is greater or smaller than
$\overline g_{\mathrm{cr}}^2$, $a$ is positive or negative. On the
other hand, the parameter $d$ [Eq.~(\rf{4e5})] is positive and
in general, for physical applications, smaller than 1; the effective
range $r_e^{}$ is then predicted positive and small.
\par
Of particular interest is the case of resonances, which appear as
bumps in the cross section above the two-particle threshold. They
correspond to complex poles of the scattering amplitude, lying below
the cut of the real axis. 
To check the possible presence of complex poles, we go back to
the expression (\rf{3e6}) of $Q(s)$ in the complex plane, which
can also be rewritten in the following form:
\be \lb{4e6}
Q(s)=\frac{\sqrt{\lambda(s)}}{s}\,\Big[-i\pi+\ln\Big
(\frac{\sqrt{s-(m_1^{}-m_2^{})^2}
+\sqrt{s-(m_1^{}+m_2^{})^2}}{\sqrt{s-(m_1^{}-m_2^{})^2}
-\sqrt{s-(m_1^{}+m_2^{})^2}}\Big)\Big].
\ee
Using a definition of the type of (\rf{3e10}), one has in approximate
form, neglecting quadratic terms in $E$,
\be \lb{4e7}
s-(m_1^{}+m_2^{})^2\simeq 2(m_1^{}+m_2^{})E,\ \ \ \ \ \
s-(m_1^{}-m_2^{})^2\simeq 4m_1^{}m_2^{}(1+\frac{E}{2m_r^{}}).
\ee
The first equation  shows that the complex variable $E$, considered
as a vector, is parallel to $(s-(m_1^{}+m_2^{})^2)$ in the $s$-plane
and therefore one can transpose to $E$ the complex-plane analysis,
with a right-hand cut starting at $E=0$.
Making expansions in $\sqrt{e}$ and retaining terms
up to order $e$, one obtains
\be \lb{4e8}
Q(s)=\alpha\sqrt{e}\,(-i\pi+2\sqrt{e}).
\ee
($\alpha$ defined in Eq.~(\rf{3e11}).)
$J(s)$ [Eq.~(\rf{3e8})] then takes the form
\be \lb{4e9}
J(s)=\frac{i}{16\pi^2}\alpha\sqrt{e}\,\Big[-i\pi+
2\sqrt{e}(1-d)\Big],
\ee
where $d$ has been defined in (\rf{4e5}). In the nonrelativistic
domain, the second term is negligible in front of the first and 
the resonance equation takes the form
\be \lb{4e10}
-i\sqrt{e_R^{}}+\frac{16\pi}{\alpha h}=0,
\ee
which yields a purely imaginary solution for $\sqrt{e_R^{}}$
and gives back the bound state solution (\rf{3e11}). Formal
resonance solutions can be obtained outside the nonrelativistic
domain for small negative values of $h$
[$-\frac{\alpha h}{16\pi}<\frac{8}{\pi}(1-d)$], by including also
the second term  of the right-hand-side of (\rf{4e9}); however,
such values of $h$ correspond, according to the correspondence
(\rf{3e12}), to the strong-coupling limit of $\overline{g}^2$,
which goes beyond the validity of the present nonrelativistic
approximation. 
\par
The previous results mean that the present model does not produce
resonances in the vicinity of the two-particle threshold. 
Resonances can be produced when there are deri\-vative-type couplings
\cite{Kang:2016ezb,Pelaez:2015qba,Pelaez:2021dak},
which we have discarded in the present
approach. One can also refer to the rectangular well model
of \cite{Nussenzveig:1959}, where resonances in the $S$-wave are
generally produced far from the real axis.
\par

\section{Compact tetraquarks} \lb{s5}

We have considered in the previous sections, in an effective theory
approach, the bound state formation problem of a molecular state, or
a hadronic molecule, which we also called a molecular-type tetraquark,
from two mesons, interacting by short-range Yukawa-type forces,
approximated in the effective theory by a contact-type interaction.
We have noticed that the smallness of the binding energy is sharpened
when the coupling constant of the lower-energy theory takes large
negative values, corresponding in the higher-energy theory to the
proximity of the coupling constant to the critical value, below which
no bound state exists.
\par
The latter mechanism is not the only one that may produce
tetraquark states. Another mechanism, based on the direct internal
interaction of four-quark systems (more precisely, made of two quarks
and two antiquarks) by means of the confining forces, might also
produce bound states, in analogy to what happens with the formation
of ordinary hadrons \cite{Maiani:2004vq,Maiani:2005pe}. Because of
the strong nature of the confining forces, one expects that such
bound states would have more compact sizes than the molecular-type
bound states and are distinguished from the latter in the literature
under the terms of ``compact tetraquarks''.
\par
However, the formation of compact tetraquarks as definite stable
bound states (with respect to the strong interactions) remains
a matter of debate. This is related to the ``cluster reducibility''
problem, in the sense that the multiquark operators that create
tetraquarks are reducible to a combination of mesonic clusters,
and hence the compact tetraquark state would rapidly dislocate
into them and would be transformed into a molecular-type object
\cite{Jaffe:2008zz,Weinstein:1982gc,Wang:1992wi,Lucha:2019cdc}.
\par
Another argument which is advocated in favor of the molecular scheme
is the proximity of many of the observed tetraquark candidate states
to two-meson thresholds. In the molecolar scheme, the two-meson
threshold is a natural reference of energy levels. In the compact
tetraquark scheme, the elementary confining forces do not refer to
meson states and hence, at first sight, no natural justification is
proposed for the appearance of tetraquark states near two-meson
thresholds.
\par
We shall analyze, in the present section, these questions with the
aid of the effective field theory approach, adopted in Secs.
\rf{s3} and \rf{s4}.
\par

\subsection{Compositeness} \lb{s51}

The comparison of the molecular and compact schemes is reminiscent
of a general problem, already raised in the past in the case of the
deuteron state, denoted under the term of ``compositeness''
\cite{Weinberg:1965zz}. The binding energy of the deuteron, referred
to the proton-neutron threshold, is very small as compared to the mass
scale involved in the strong interaction dynamics of the nucleons.
One is inclined to consider the deuteron as a loosely bound composite
object, or a molecule, made of a neutron and a proton. On the other
hand, there might still exist some probability, that should be
quantified, that it might be an elementary particle, or a compact
object. Weinberg has shown that this question can receive, in the
nonrelativistic limit, a precise and model-independent answer, by
relating the latter probability to observable quantities, represented
by the scattering length and the effective range of the neutron-proton
$S$-wave isospin-0 scattering amplitude \cite{Weinberg:1965zz}.
Designating by $Z$ the probability of finding the deuteron in an
elementary, or compact, state, Weinberg has found the following
relations for the scattering length $a$ and the effective range
$r_e^{}$:
\be \lb{5e1}
a=\frac{2(1-Z)}{(2-Z)}R+O(m_{\pi}^{-1}),\ \ \ \ \ 
r_e^{}=-\frac{Z}{(1-Z)}R+O(m_{\pi}^{-1}),\ \ \ \ \ 
R=(-2m_r^{}E_d^{})^{-1/2},
\ee
where $R$ is the deuteron radius, $m_r^{}$ the reduced mass of the
proton-neutron system [Eq.~(\rf{2e3})], $E_d^{}$
($\equiv 2m_r^{}e_d^{}$) the deuteron
nonrelativistic energy (opposite of its binding energy) and
$m_{\pi}^{}$ the pion mass; $O(m_{\pi}^{-1})$ represents the scale
of the hadronic corrections that are negligible in front of $R$. On
the other hand, introducing the dimensionless
deuteron-neutron-proton coupling constant $g_{dnp}^{}$, accompanied
by the factor $(m_n^{}+m_p^{})$, one has the relationship
\be \lb{5e2}
g_{dnp}^2=32\pi\sqrt{-e_d}\ (1-Z).
\ee
\par
One notices that the effective range is the most sensitive quantity
to $Z$, which, in case $Z\neq 0$, is manifested by a sizeable
negative value.
Using Eqs. (\rf{5e1}), one can also express the compositeness
factor in a combined form with respect to $a$ and $r_e^{}$:
\be \lb{5e2a}
1-Z=\frac{1}{\sqrt{1-2r_e^{}/a}}.
\ee
\par
In the case of the deuteron, the experimental data
about $a$ and $r_e^{}$ rule out a nonzero value of $Z$
and confirm its composite nature \cite{Weinberg:1965zz}.
\par
\par
Equations (\rf{5e1}) and (\rf{5e2}) can also be used, with
appropriate relabeling of the parameters, to check the consistency
of the results obtained in Secs. \rf{s3} and \rf{s4}.
Equations (\rf{4e4}) show that $r_e^{}$ has a small negligible 
value (as compared to $R=1/(2m_r^{}\sqrt{-e_{tm}^{}})$), which could
be interpreted as representing the higher-order hadronic corrections.
Thus, with respect to the second of Eqs. (\rf{5e1}), the main
value of $r_e^{}$ is $0$, which entails that $Z=0$, in accordance
with the molecular nature of the bound state. Also, the comparison
of Eq.~(\rf{3e16}) with (\rf{5e2}) confirms the latter
conclusion.
\par
For later reference, using notations adapted to the tetraquark
problem, we display here the expression of the scattering
amplitude obtained in Weinberg's analysis:
\be \lb{5e2b}
\mathcal{T}=8\pi(m_1^{}+m_2^{})
\Big[\,32\pi\,\frac{(e_t^{}-e)}{g_{TM_1^{}M_2^{}}^2}\,m_r^{}
+\frac{(e_t^{}+e)}{\sqrt{-e_t^{}}}\,m_r^{}-ik\,\Big]^{-1},
\ee
where $g_{TM_1^{}M_2^{}}^{}$ has been defined in (\rf{3e14})
and $e$ and $e_t^{}$ in (\rf{3e10a}). The main assumption
underlying this result concerns the absence of zeros in
$\mathcal{T}$, at least in the vicinity of the bound state
\cite{Castillejo:1955ed}.
\par

\subsection{Compact bound states} \lb{s52}

We consider, in this subsection, the case of possibly existing
compact tetraquarks.
We shall not enter, for the analysis of the problem,  into the
details of the mechanism that produces such states, but merely
shall assume their existence.
If experimental data were sufficiently
precise concerning an observed tetraquark candidate, providing us
with its coupling amplitude to the nearby two-meson states, as well
as the scattering length and the effective range of the related
two-meson elastic scattering amplitude, then, for a
non\-rela\-ti\-vis\-tic state, Eqs. (\rf{5e1}) and (\rf{5e2})
would allow us to reach a conclusion about the internal
structure of the tetraquark. In the absence of high precision
data, we proceed by successive steps. 
In first approximation, we assume that the compact tetraquark is
a pointlike object in comparison to a loosely bound molecular state.
At this stage, we assume that the mass of the tetraquark has been
evaluated by the sole mechanism of the confining forces, from which
clustering forces or effects have been removed.
(The small volume or pointlike approximations of the diquark system
satisfy this requirement.)
Furthermore, we assume that the mass of the bound state under study,
designated by $m_{tc1}^{}$, where the labels $tc$ refer to the
compact tetraquark, is rather close to the nearest two-meson
threshold mass $(m_1^{}+m_2^{})$ and, therefore, a nonrelativistic
energy of the bound state, $E_{tc1}^{}$, can be defined by means of
the equation
\be \lb{5e3}
E_{tc1}^{}=m_{tc1}^{}-(m_1^{}+m_2^{}),
\ee
$E_{tc1}^{}$ remaining a small quantity with respect to the two-meson
reduced mass. However, we do not assume that the bound state is
shallow. Considering the case of the deuteron as an example of a
shallow bound state, whose binding energy is of the order
of 2 MeV, $E_{tc1}^{}$ might have values of the order of 20-30 MeV.
\par
Another point worth emphasizing is that, in general, when one has
many different quark flavors inside the tetraquark state, the latter
has two different two-meson clusters
\cite{Lucha:2017mof,Lucha:2017gqq,Lucha:2021mwx} and one should,
in that case, use
a coupled-channel formalism. However, in order to display in a
clearer way the main qualitative aspects of the problem, we stick
here to a single-channel formalism (which describes an exact
situation when two quarks, or two antiquarks, have the same flavor).
\par
Because of the existence of internal two-meson clusters inside the
tetraquarks, the compact tetraquark has necessarily a coupling to
the two mesons $M_1^{}$ and $M_2^{}$. The corresponding
(dimensionless) coupling constant is designated by $g'$, factored by
the mass term $(m_1^{}+m_2^{})$. However, the latter coupling
generates, through meson loops, radiative corrections inside the
tetraquark propagator, thus modifying the parameters of the bare
propagator. They are graphically represented in Fig.~\rf{f5}.  
\par
\bfg
\vspace*{0.25 cm}
\bc
\epsfig{file=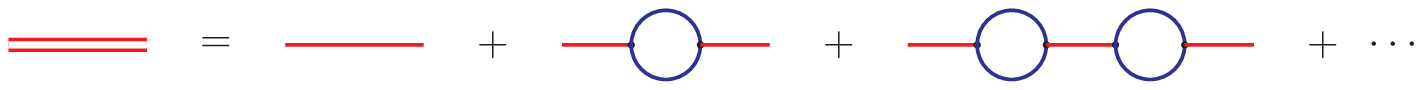,scale=0.8}
\vspace*{0.25 cm}
\caption{Chain of meson-one-loop radiative corrections to the
tetraquark propagator.} 
\lb{f5}
\ec
\efg
\par
Designating by $m_{tc0}^{}$ the bare tetraquark mass, the full
tetraquark propagator becomes
\be \lb{5e4}
D_{tc}^{}(s)=\frac{i}{s-m_{tc0}^2+i(m_1^{}+m_2^{})^2g^{\prime 2}J(s)},
\ee
where $s$ stands for $p^2$ and $J$ is the same loop function as
the one met in Eqs. (\rf{3e2})-(\rf{3e8}). The divergence of
$J$ is now absorbed by the bare mass term, yielding the renormalized
mass $m_{tc1}^{}$:
\be \lb{5e5}
m_{tc1}^2=m_{tc0}^2-i (m_1^{}+m_2^{})^2g^{\prime 2}J^{\mathrm{div}}.
\ee
The renormalized tetraquark propagator is now
\be \lb{5e6}
D_{tc}^{}(s)=\frac{i}{s-m_{tc1}^2+i(m_1^{}+m_2^{})^2g^{\prime 2}
J^{\mathrm{r}}(s)}.
\ee
Notice that $g'$ does not undergo any renormalization. The mass
term $m_{tc1}^{}$ does not yet represent the physical mass of the
tetraquark. The latter is determined from the pole position of the
propagator. Sticking to the nonrelativistic limit, one can use
for $s$ and $m_{tc1}^{}$ expansions of the types of (\rf{3e10})
and (\rf{5e3}). Retaining in $J^{\mathrm{r}}(s)$ the dominant
contribution, one finds for the nonrelativistic energy of the
tetraquark, designated by $E_{tc}^{}$, the equation
\be \lb{5e7}
-e_{tc}^{}+e_{tc1}^{}+\frac{g^{\prime 2}}{16\pi}\sqrt{-e_{tc}^{}}=0,
\ee
whose solution is
\be \lb{5e8}
\sqrt{-e_{tc}^{}}=\frac{1}{2}\Big[-\frac{g^{\prime 2}}{16\pi}+
\sqrt{\Big(\frac{g^{\prime 2}}{16\pi}\Big)^2-4e_{tc1}^{}}\Big].
\ee
The binding energy $(-e_{tc}^{})$, being a decreasing function of
$g^{\prime 2}/(16\pi)$, comes out, in general, smaller than
$(-e_{tc1}^{})$, reaching the value 0 when $g'\rightarrow\infty$
(see Fig.~\rf{f6}).
\bfg
\vspace*{0.25 cm}
\bc
\epsfig{file=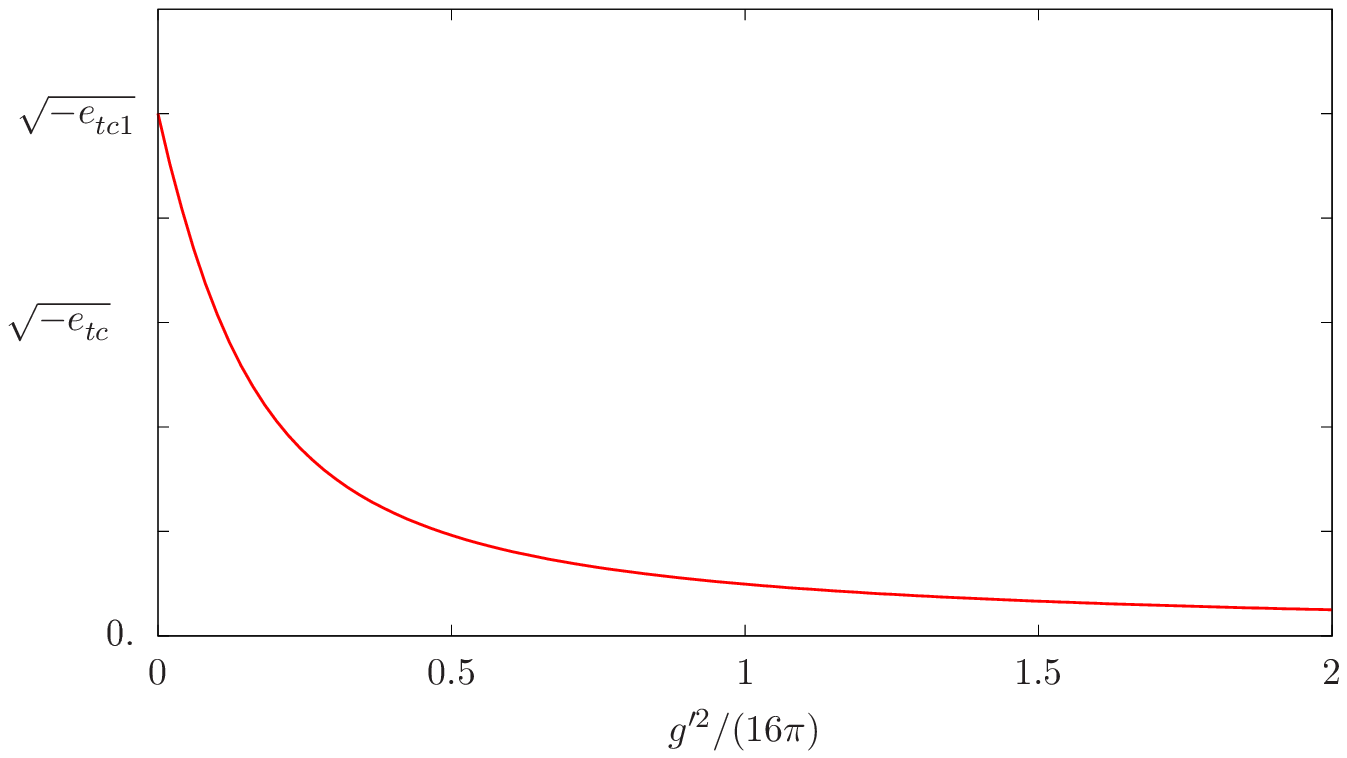,scale=0.8}
\vspace*{0.25 cm}
\caption{Variation of the square-root of the binding energy as a
function of $g^{\prime 2}/(16\pi)$. The value of
$-e_{tc1}$ $(\equiv -E_{tc1}^{}/(2m_r^{}))$ has been fixed at 0.01.}  
\lb{f6}
\ec
\efg
\par
Of particular interest are the weak- and strong-coupling limits in
$g'$, determined by the comparison of the factors
$g^{\prime 2}/(16\pi)$ and $\sqrt{-4e_{tc1}^{}}$.
In the first case,
$g^{\prime 2}/(16\pi)\ll\sqrt{-4e_{tc1}^{}}$, $-e_{tc}^{}$ is
obtained close to $-e_{tc1}^{}$ with a small negative shift. In the
second case, $g^{\prime 2}/(16\pi)\gg\sqrt{-4e_{tc1}^{}}$,
one obtains
\be \lb{5e9}
\sqrt{-e_{tc}^{}}\simeq -e_{tc1}^{}\,\frac{16\pi}{g^{\prime 2}}.
\ee
(This expression could also be obtained directly from (\rf{5e7})
by neglecting in it $e_{tc}^{}$ in front of $e_{tc1}^{}$.)
Because of the nonlinear relationship between $e_{tc}^{}$ and
$e_{tc1}^{}$, there appears a strong decrease in $e_{tc}^{}$. One has
\be \lb{5e10}
\frac{e_{tc}^{}}{e_{tc1}^{}}=-e_{tc1}^{}\,
\Big(\frac{16\pi}{g^{\prime 2}}\Big)^2.
\ee
Considering for example $-E_{tc1}^{}\simeq 20$ MeV and
$2m_r^{}\simeq 2$ GeV, one has $-E_{tc1}^{}/(2m_r^{})\simeq 0.01$;
values of $g^{\prime 2}/(16\pi)$ of the order of or greater than
0.5 produce ratios $E_{tc}^{}/E_{tc1}^{}\leq 0.04$, or equivalently
$-E_{tc}^{}\leq 0.8$ MeV. These values of $g^{\prime 2}/(16\pi)$ are
not exceptional and we may conclude that we are not in the presence
of a fine tuning effect. There is a nonnegligible probability, in
many physical cases, to meet such a situation.
\par
The contribution of the tetraquark state, in the $s$-channel, to
the two-meson elastic scattering amplitude is obtained by inserting
the tetraquark propagator between two tetraquark-two-meson couplings,
as shown in Fig.~\rf{f7}.
\bfg
\vspace*{0.25 cm}
\bc
\epsfig{file=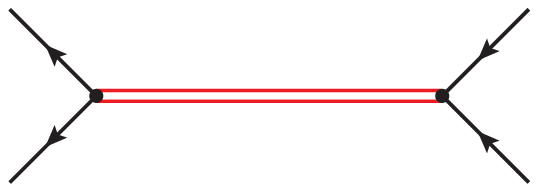,scale=0.8}
\vspace*{0.25 cm}
\caption{The tetraquark contribution, in the $s$-channel, to the
two-meson elastic scattering amplitude.}
\lb{f7}
\ec
\efg
\par
One finds
\be \lb{5e11}
\mathcal{T}=-\frac{(m_1^{}+m_2^{})^2g^{\prime 2}}
{s-m_{tc1}^2+i(m_1^{}+m_2^{})^2g^{\prime 2}
J^{\mathrm{r}}(s)}.
\ee
Proceeding as in the molecular case [Eq.~(\rf{3e14})], one can
obtain the physical coupling constant $g_{TM_1^{}M_2^{}}^{}$ of the
tetraquark to the two mesons:
\be \lb{5e12}
g_{TM_1^{}M_2^{}}^2=32\pi\sqrt{-e_{tc}^{}}\
\frac{1}{\Big[1+2\Big(\frac{16\pi}{g^{\prime 2}}\Big)^2
(e_{tc}^{}-e_{tc1}^{})\Big]}.
\ee
Comparing this expression with (\rf{5e2}), one obtains $Z$:
\be \lb{5e13}
Z=\frac{1}{1+\Big(\frac{g^{\prime 2}}{16\pi}\Big)^2
\frac{1}{2(e_{tc}^{}-e_{tc1}^{})}},
\ee
which, after taking into account Eq.~(\rf{5e7}), can also
be expressed as
\be \lb{5e14}
Z=\frac{\sqrt{-e_{tc}^{}}}{\sqrt{-e_{tc}^{}}+\frac{1}{2}
\frac{g^{\prime 2}}{16\pi}}.
\ee
One notices, in particular from (\rf{5e13}), that $Z$ decreases in
the strong-coupling limit and takes small values. With the numerical
example considered above, one has $Z\leq 0.075$. On the other hand, in
the same limit, the physical coupling constant $g_{TM_1^{}M_2^{}}^{}$
also decreases, like $\sqrt{-e_{tc}^{}}$. The variation of $Z$ as
a function of $g^{\prime 2}/(16\pi)$, taking into account (\rf{5e8}),
is represented in Fig.~\rf{f8}. One notices the rapid decrease of
$Z$.
\bfg
\vspace*{0.25 cm}
\bc
\epsfig{file=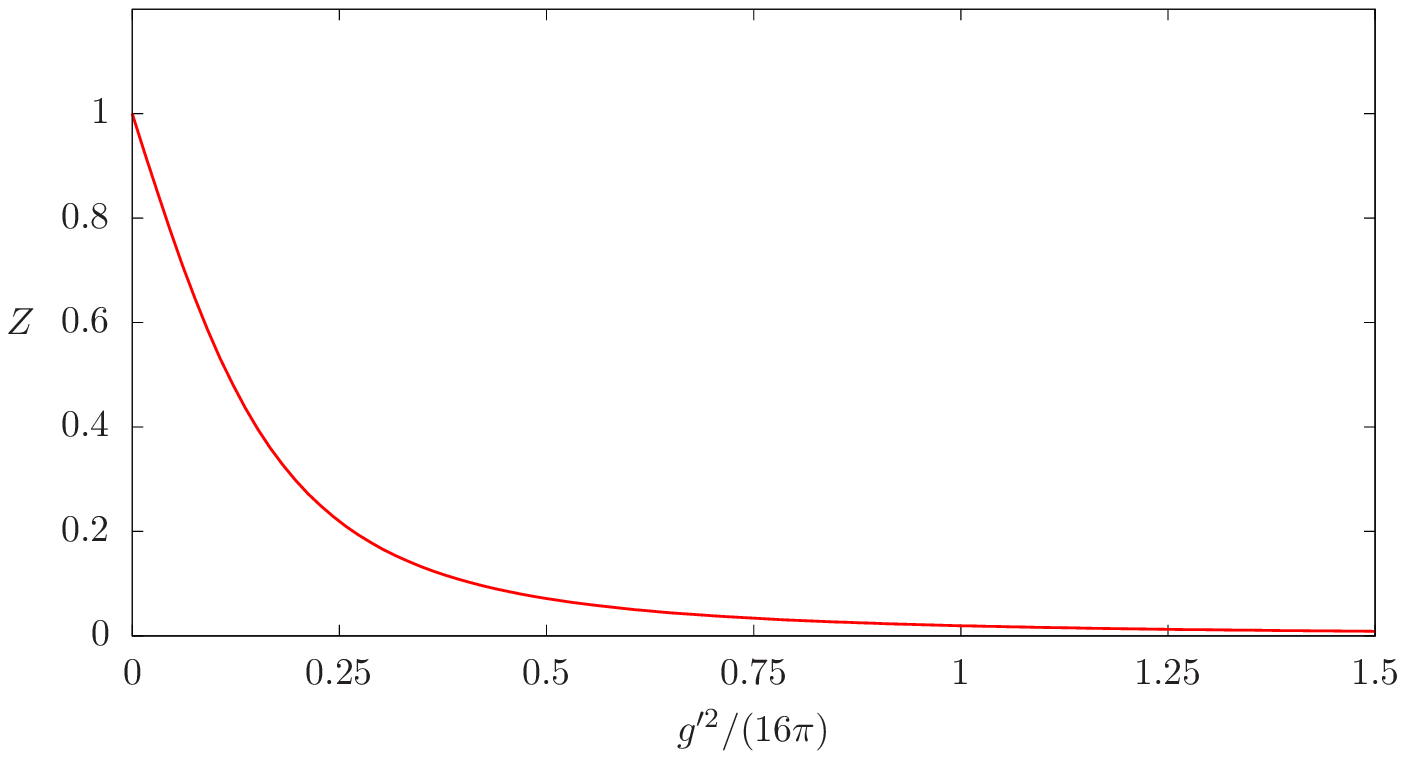,scale=0.8}
\vspace*{0.25 cm}
\caption{Variation of $Z$ as a function of $g^{\prime 2}/(16\pi)$;
$-e_{tc1}^{}$ has been fixed at 0.01.}
\lb{f8}
\ec
\efg
\par
It is also possible to calculate from (\rf{5e11}), as in the molecular
case [Eqs. (\rf{4e1})-(\rf{4e5})], the scattering length and the
effective range. One finds, neglecting other contributions to the
scattering amplitude,
\be \lb{5e15}
a=\Big(\frac{g^{\prime 2}}{16\pi}\Big)\,
\Big(\frac{1}{2m_r^{}(-e_{tc1}^{})}\Big),\ \ \ \ \ \ \
r_e^{}=\frac{1}{m_r^{}}\Big(-\frac{16\pi}{g^{\prime 2}}+
\frac{2}{\pi}(1-d)\Big).
\ee
($d$ is defined in (\rf{4e5}).) Eliminating $g^{\prime 2}/(16\pi)$
and $e_{tc1}^{}$ in favor of $Z$ and $e_{tc}^{}$, one obtains the
expected expressions
\be \lb{5e16}
a=\frac{2(1-Z)}{(2-Z)}R,\ \ \ \ \ \ \ 
r_e^{}=-\frac{Z}{(1-Z)}R,\ \ \ \ \ \ \
R=\frac{1}{2m_r^{}}\sqrt{\frac{1}{-e_{tc}^{}}}.
\ee
(The second term in $r_e^{}$, which is small, has been neglected.)
\par
It is of importance to have a precise interpretation of the values
of $Z$. When $g'=0$, $Z=1$ and the tetraquark is completely decoupled
from the two mesons. This would mean that either its mass scale or 
quark content are different from those of the two mesons. In the
opposite case, corresponding to the strong-coupling limit, $Z$
decreases and approaches the value 0.
This does not mean, however, that the tetraquark's nature becomes
molecular. Nowhere, in the present model, did we
consider direct interactions between mesons; the existence of the
bound state is entirely due to the confining forces that are
responsible of its compact nature. The value of $(1-Z)$ simply
reflects the strength of the (bare, but finite) coupling of the
tetraquark to the two meson clusters. As $g'$ grows, the latter,
through the radiative corrections, plays, in the internal structure
of the tetraquark, an increasingly determinant role, leading to a
strong decrease of the binding energy and to a corresponding
increase of the radius $R$ [Eq.~(\rf{5e16})]. The tetraquark,
though of compact nature, is gradually deformed into a
molecular-type state.
\par

\subsection{Resonances} \lb{s53}

A resonance may occur when the renorma\-lized (real) mass of the
com\-pact tetra\-quark [Eq.~(\rf{5e5})] lies above the two-meson
threshold. Its nonrelativistic energy, defined in (\rf{5e3}), is
now positive. The position of the complex mass of the possibly
existing resonance can be searched for with the same method and the
same approximations as in the bound state case. Designating by
$E_{TR}^{}$ the nonrelativistic (complex) energy of the tetraquark
resonance, the equivalent of Eq.~(\rf{5e7}) is
\be \lb{5e17}
(e_{TR}^{}-e_{tc1}^{})+i\frac{g^{\prime 2}}{16\pi}
\sqrt{e_{TR}^{}}=0.
\ee
Its solutions are
\be \lb{5e18}
\sqrt{e_{TR}^{}}=-\frac{i}{2}\frac{g^{\prime 2}}{16\pi}\pm
\sqrt{e_{tc1}^{}
-\frac{1}{4}\Big(\frac{g^{\prime 2}}{16\pi}\Big)^2}.
\ee
The imaginary part of $\sqrt{e_{TR}^{}}$
comes out negative, which means that the two solutions lie in the
second Riemann sheet. The expression of $e_{TR}^{}$ is
\be \lb{5e19}
e_{TR}^{}=\Big[e_{tc1}^{}-\frac{1}{2}
\Big(\frac{g^{\prime 2}}{16\pi}\Big)^2\Big]\mp 
i\frac{g^{\prime 2}}{16\pi}
\sqrt{e_{tc1}^{}-\frac{1}{4}\Big(\frac{g^{\prime 2}}{16\pi}\Big)^2}.
\ee
The two solutions are complex conjugate to each other, the resonance
corresponding to the negative imaginary part.
One verifies that the modulus of $e_{TR}^{}$ is equal to $e_{tc1}^{}$:
\be \lb{5e20a}
|e_{TR}^{}|=e_{tc1}^{}.
\ee
The condition of the positivity of the real part of $e_{TR}^{}$
(resonance above the threshold) requires that
\be \lb{5e20}
\Big(\frac{g^{\prime 2}}{16\pi}\Big)^2\leq 2\,e_{tc1}^{}.
\ee
This means that one is rather in the weak-coupling regime. As the
coupling constant increases, the real part of the resonance energy
approaches the threshold, the upper bound of (\rf{5e20})
corresponding to its merging with the threshold. On the other hand,
the imaginary part remains always different from zero; at threshold,
it is only the latter that survives.
\par
The scattering amplitude, due to the resonance, is
\be \lb{5e21}
\mathcal{T}=-\frac{(m_1^{}+m_2^{})}{4m_r^{}}\
\frac{g^{\prime 2}}{\Big(e-e_{tc1}^{}
+i\,\frac{g^{\prime 2}}{16\pi}\sqrt{e}\Big)}.
\ee
We have represented, in Fig.~\rf{f9}, the shape of
$|\mathcal{T}|^2$ in the vicinity of the threshold, for real $E$ and
for several values of $g^{\prime 2}/(16\pi)$ in its allowed domain.
\bfg
\vspace*{0.25 cm}
\bc
\epsfig{file=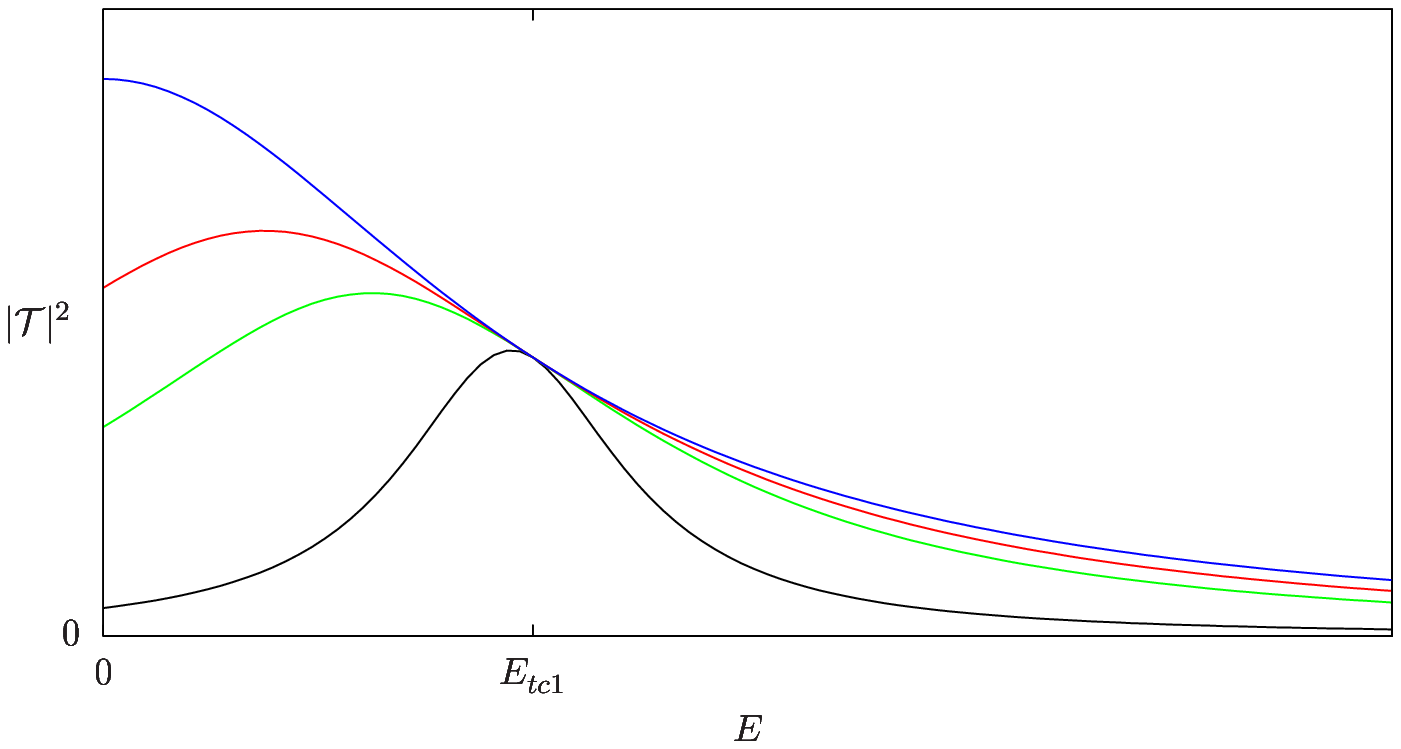,scale=0.8}
\vspace*{0.25 cm}
\caption{The shape of $|\mathcal{T}|^2$ for several values of
$g^{\prime 2}/(16\pi)$: $\Big(g^{\prime 2}/(16\pi)\Big)^2/
e_{tc1}^{}=$ 0.1 (black curve), 0.75 (green), 1.25 (red),
2.0 (blue).}  
\lb{f9}
\ec
\efg
\par
Expanding $\mathcal{T}$ around $E_{TR}^{}$ [Eq.~(\rf{5e19})]
and using for the resonance the plus sign in (\rf{5e18}), one
obtains
\be \lb{5e22}
\mathcal{T}\simeq -\frac{(m_1^{}+m_2^{})g_{TM_1^{}M_2^{}}^2}
{2(E-E_{TR}^{})},
\ee
with
\be \lb{5e23}
g_{TM_1^{}M_2^{}}^2=\frac{32\pi\sqrt{e_{TR}^{}}}
{\sqrt{4e_{tc1}^{}\Big(\frac{16\pi}{g^{\prime 2}}\Big)^2-1}}.
\ee
Comparing (\rf{5e23}) with (\rf{5e2}), the former being considered
as a formal extension of the latter to the resonance region, one
obtains the compositeness coefficient
\be \lb{5e24}
1-Z=\frac{1}{\sqrt{4e_{tc1}^{}
\Big(\frac{16\pi}{g^{\prime 2}}\Big)^2-1}}.
\ee
We notice that, due to the inequality (\rf{5e20}), which guarantees
the occurrence of the resonance above the threshold, 
$(1-Z)$ is real, positive and smaller than 1. This allows us to
continue giving it a probabilistic interpretation.
\par
Writing the (reduced) energy of the resonance in the form
\be \lb{5e25}
e_{TR}^{}=e_{TRr}^{}-i\frac{\gamma}{2},
\ee
where $e_{TRr}^{}$ is the real part of the (reduced) energy and
$\gamma$ the reduced dimensionless width of the resonance,
defined as (see also (\rf{3e10a}))
\be \lb{5e26}
\gamma\equiv\frac{\Gamma}{2m_r^{}},
\ee
$\Gamma$ being the dimensionful width, one has from (\rf{5e19})
\be \lb{5e28}
e_{TRr}^{}=e_{tc1}^{}-\frac{1}{2}\Big(\frac{g^{\prime 2}}{16\pi}\Big)^2,
\ \ \ \ \ \ \
\gamma=2\frac{g^{\prime 2}}{16\pi}\sqrt{e_{tc1}^{}
-\frac{1}{4}\Big(\frac{g^{\prime 2}}{16\pi}\Big)^2},
\ee
which allows us to express $(1-Z)$ in terms of observable
quantities\footnote{The following relations are also useful:
$\sqrt{4e_{TRr}^2/\gamma^2+1}=((1-Z)^{-1}+(1-Z))/2$ and
$2e_{TRr}^{}/\gamma=((1-Z)^{-1}-(1-Z))/2$.}:
\be \lb{5e29}
1-Z=\frac{\Gamma/2}{\Big(E_{TRr}^{}
+\sqrt{E_{TRr}^2+\frac{\Gamma^2}{4}}\Big)}.
\ee
\par
Of particular interest is the case of narrow resonances,
characterized by the inequality $\Gamma\ll 2E_{TRr}^{}$, which
entails
\be \lb{5e30}
1-Z\simeq \frac{\Gamma}{4E_{TRr}^{}},\ \ \ \ \ \ \
\Gamma\ll 2E_{TRr}^{}.
\ee
It is evident from this result that the narrow resonance case favors
values of $Z$ close to 1, that is, dominance of the compact nature
of the tetraquark. In the opposite case, when the inequality 
(\rf{5e20}) is saturated, one has vanishing of $e_{TRr}^{}$ and
merging of the resonance with the threshold, with $Z=0$; the trace
of the compact origin of the resonance is then completely lost.
\par
The expressions of the scattering length and the effective range
are the same as in Eqs. (\rf{5e15}), except that $e_{tc1}^{}$
is now positive:
\bea
\lb{5e31}
& &a=-\frac{g^{\prime 2}}{16\pi}\,\frac{1}{2m_r^{}e_{tc1}^{}}
=-\frac{1}{2m_r^{}}\frac{\gamma/\sqrt{2}}
{\sqrt{e_{TRr}^2+\frac{\gamma^2}{4}}\sqrt{e_{TRr}^{}
+\sqrt{e_{TRr}^2+\frac{\gamma^2}{4}}}},\\    
\lb{5e32}
& &r_e^{}=-\frac{16\pi}{g^{\prime 2}}\,\frac{1}{m_{r}^{}}
=-\frac{1}{m_r^{}\gamma/\sqrt{2}}\sqrt{e_{TRr}^{}
+\sqrt{e_{TRr}^2+\frac{\gamma^2}{4}}}.
\eea
(The second term in $r_e^{}$, which is small, has been neglected.)
One notices that now, as compared to the bound state case
(\rf{5e15}), both the scattering length and the effective range
are negative. In terms of $a$ and $r_e^{}$, the compositeness
coefficient (\rf{5e24}) takes the form
\be \lb{5e33}
1-Z=\frac{1}{\sqrt{\frac{2r_e^{}}{a}-1}},
\ee
a relation also obtained in \cite{Kang:2016ezb}. It can be compared
with Eq.~(\rf{5e2a}), valid in the bound state case.
In the narrow resonance case, one has the simplified expressions
\be \lb{5e34}
a\simeq -\frac{1}{4m_r^{}}\frac{\gamma}{e_{TRr}^{3/2}},\ \ \ \ \ \
r_e^{}\simeq -\frac{2}{m_r^{}}\frac{e_{TRr}^{1/2}}{\gamma}.
\ee
\par
Let us also comment on the case of the strong-coupling limit,
where, in particular, the inequality (\rf{5e20}) is no longer
satisfied. We have to distinguish here two cases. The first
corresponds to the domain
$2e_{tc1}^{}<(g^{\prime 2}/(16\pi))^2<4e_{tc1}^{}$, in which case
$(1-Z)>1$ and the real part of the energy becomes negative
[Eq. (\rf{5e28})]. This is a sign of the instability of the
initial system under the influence of the two-meson clusters and
might signify the disappearance of the compact tetraquark from
the spectrum. The second corresponds to the domain
$4e_{tc1}^{}<(g^{\prime 2}/(16\pi))^2$. In that case,
$\sqrt{e_{TR}^{}}$ [Eq.~(\rf{5e18})] becomes imaginary and
$e_{TR}^{}$ becomes real and negative, falling back in the
bound-state regime. Considering then the new value of $e_{TR}^{}$
as a starting value (equal to a new $e_{tc1}^{}$), one continues
remaining in the bound-state regime for any value of the coupling
constant, as we have seen in Eq.~(\rf{5e8}) and Fig.~\rf{f6}.
Therefore, genuine resonances may occur, in the present model, only
in the weak-coupling regime, satisfying the inequality (\rf{5e20}).
\par
Let us notice, as a final remark, that due to the fact that the
internal structure of the compact elementary particle, assumed
here as being a tetraquark, was not specified, one is also entitled
to apply the previous approach, when the flavor and other quantum
numbers are compatible, to ordinary mesons. In that case, its field
theoretic basis is even more robust. This is supported by the
large-$N_c^{}$ limit of QCD \cite{'tHooft:1973jz,Witten:1979kh,
Coleman:1985rnk,Lucha:2021mwx}. In that limit, the spectrum of the
theory is composed of free mesons, which are made of one
quark-antiquark pair. They do not have any internal other meson
clusters. Therefore, their elementary nature with respect to the
other mesons is well justified. The couplings to the other mesons
appear only at nonleading order in $N_c^{}$, putting
the clustering phenomenon at a perturbative level. This is in
contrast to the multiquark case, where the clustering occurs
already at leading order in $N_c^{}$ and becomes even stronger
in the large-$N_c^{}$ limit \cite{Lucha:2021mwx}.
(Cf. also Sec.~\rf{s7}.)
\par
Detailed investigations about the compositeness criterion and its
applicability to various tetraquark candidates, as well as to
ordinary hadrons, can be found in
References \cite{Baru:2003qq,Cleven:2011gp,Hanhart:2011jz,
Hyodo:2011qc,Aceti:2012dd,Sekihara:2014kya,Guo:2015daa,Kang:2016ezb,
Meissner:2015mza,Oller:2017alp,Guo:2020vmu,Esposito:2021vhu,
Li:2021cue,Baru:2021ldu,Kinugawa:2021ykv,Song:2022yvz}.
\par

\section{Presence of meson-meson interactions} \lb{s6}

In the model considered in Sec. \rf{s5}, where the influence of
the coupling of a compact tetraquark to two mesons was studied,
the presence of meson-meson interactions as a background effect
was not taken into account. This had the advantage of exhibiting
in a clearer way the role of the aforementioned coupling on the
observable properties of the tetraquark state, in particular its
gradual deformation, in the strong coupling limit, towards a
molecular-type object. To complete the previous study, we include,
in this section, the effect of the meson-meson interaction into
the dynamical process.
\par

\subsection{The meson-meson scattering amplitude} \lb{s61}

The meson-meson interaction, in the present effective field theory
description, was considered in Secs. \rf{s3} and \rf{s4}. When
a compact tetraquark state is present, its effect is first manifested
through a vertex renormalization related to the coupling constant
$g'$. This is graphically represented in Fig.~\rf{f10}.
\bfg
\vspace*{0.25 cm}
\bc
\epsfig{file=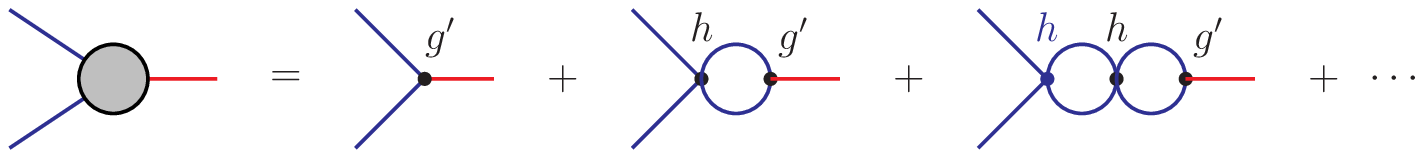,scale=0.8}
\vspace*{0.25 cm}
\caption{Chain of meson-one-loop radiative corrections to the
tetraquark-two-meson coupling constant $g'$. The coupling constants
at the vertices are indicated. $g'$ is accompanied by the mass
factor $(m_1^{}+m_2^{})$.} 
\lb{f10}
\ec
\efg
\par
The corresponding vertex function, denoted
$\Gamma_{TM_1^{}M_2^{}}^{(3)}$,
takes the form (cf. Eqs. (\rf{3e1})-(\rf{3e4a}))
\be \lb{6e1}
\Gamma_{TM_1^{}M_2^{}}^{(3)}(s)=\frac{i(m_1^{}+m_2^{})g'}{1-ihJ(s)}.
\ee
We have seen that the divergence contained in $J(s)$
[Eq.~(\rf{3e3})] is absorbed by a renormalization of the coupling
constant $h$, according to (\rf{3e4}) or (\rf{3e4a}). Here, the
occurrence of the same divergence necessitates a similar
renormalization of $g'$, involving, however, also $h$:
\be \lb{6e2}
g'=\frac{g^{\prime \mathrm{r}}}{1+ih^{\mathrm{r}}J^{\mathrm{div}}}.
\ee
Equivalently, one has the following renormalizations:
\be \lb{6e3}
\frac{h}{1-ihJ(s)}=
\frac{h^{\mathrm{r}}}{1-ih^{\mathrm{r}}J^{\mathrm{r}}(s)},
\ \ \ \ \ \ \ \ \ \ \ \ \  
\frac{g'}{1-ihJ(s)}=
\frac{g^{\prime \mathrm{r}}}{1-ih^{\mathrm{r}}J^{\mathrm{r}}(s)}.
\ee
These ensure the renormalization of $\Gamma_{TM_1^{}M_2^{}}^{(3)}$:
\be \lb{6e1a}
\Gamma_{TM_1^{}M_2^{}}^{(3)}(s)=
\frac{i(m_1^{}+m_2^{})g^{\prime \mathrm{r}}}
{1-ih^{\mathrm{r}}J^{\mathrm{r}}(s)}.
\ee
\par
A second effect of the meson-meson interactions is manifested through
the radiative corrections of meson-one-loop diagrams, occurring in the
tetraquark propagator (cf. Fig.~\rf{f5}). Each loop receives
radiative corrections, as represented in Fig.~\rf{f11}.
\bfg
\vspace*{0.25 cm}
\bc
\epsfig{file=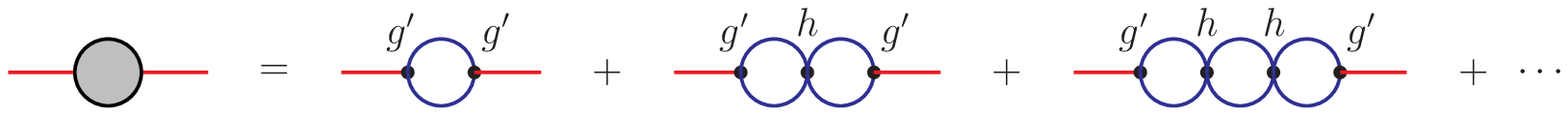,scale=0.8}
\vspace*{0.25 cm}
\caption{Chain of meson-one-loop radiative corrections to the
one-loop diagram of the tetraquark propagator. The coupling constants
at the vertices are indicated.} 
\lb{f11}
\ec
\efg
\par
The chain of such diagrams can be summed and yields the full one-loop
contribution, which we designate by $\Gamma_{TT}^{(2)}(s)$. One
obtains
\be \lb{6e4}
\Gamma_{TT}^{(2)}(s)=(m_1^{}+m_2^{})^2\frac{(ig')^2J(s)}{1-ihJ(s)}.
\ee
These full one-loop contributions replace now the simple loop
contributions of Fig.~\rf{f5}. The full tetraquark propagator is
then given by the series of diagrams of Fig.~\rf{f12}. 
\bfg
\vspace*{0.25 cm}
\bc
\epsfig{file=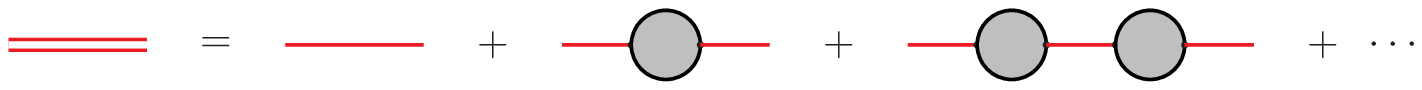,scale=0.8}
\vspace*{0.25 cm}
\caption{The full tetraquark propagator.}
\lb{f12}
\ec
\efg
\par
One finds for the full propagator
\be \lb{6e5}
D_t^{}(s)=\frac{i}{s-m_{tc0}^2+i(m_1^{}+m_2^{})^2
g^{\prime 2}J(s)/(1-ihJ(s))}.
\ee
Using the renormalizations given in (\rf{6e2}) and (\rf{6e3}),
and after taking the limit $iJ^{\mathrm{div}}\rightarrow \infty$,
one finds
\be \lb{6e6}
D_t^{}(s)=\frac{i}{s-m_{tc0}^2+(m_1^{}+m_2^{})^2
g^{\prime \mathrm{r}2}/[h^{\mathrm{r}}
(1-ih^{\mathrm{r}}J^{\mathrm{r}}(s))]}.
\ee
One notices that after the renormalizations of the coupling constants
$h$ and $g'$ have been realized, the radiative corrections of the
tetraquark propagator are now finite. This is in contrast to the
case where the meson-meson interactions had been ignored ($h=0$)
and the divergence of the radiative corrections had been absorbed by
the mass renormalization, while the coupling constant $g'$ had
remained finite (cf. (\rf{5e4}) and (\rf{5e5})). Nevertheless, the
radiative corrections in (\rf{6e6}) contain a singularity in
$h^{\mathrm{r}}$. When $h^{\mathrm{r}}\rightarrow 0$, one
recovers the divergence that exists in the aforementioned case.
To remedy this defect, one has to subtract that singularity from
the global radiative corrections and associate it with a
renormalization of the bare mass $m_{tc0}^{}$. Designating by
$m_{tc1}$ the renormalized bare mass, one has
\be \lb{6e7}
m_{tc1}^2=m_{tc0}^2-\frac{(m_1^{}+m_2^{})^2
g^{\prime \mathrm{r}2}}{h^{\mathrm{r}}}.
\ee
As long as $h^{\mathrm{r}}$ is nonzero, this mass renormalization
is finite. When $h^{\mathrm{r}}\rightarrow 0$, one recovers the
situation of Eq.~(\rf{5e5}).
\par
The full tetraquark propagator takes now the following form:
\be \lb{6e8}
D_t^{}(s)=\frac{i}{s-m_{tc1}^2+i(m_1^{}+m_2^{})^2
g^{\prime \mathrm{r}2}J^{\mathrm{r}}(s)/
(1-ih^{\mathrm{r}}J^{\mathrm{r}}(s))}.
\ee
When $h^{\mathrm{r}}\rightarrow 0$, one also recovers the propagator
(\rf{5e6}).
\par
The meson-meson scattering amplitude is obtained by inserting the
tetraquark propagator inside two vertices of the type of (\rf{6e1}),
which are finite [Eq.~(\rf{6e1a})], and by adding the
contribution generated by the contact interaction
(\rf{2e8}) [Fig.~\rf{f2} and Eqs.~(\rf{3e1}) and (\rf{3e4})].
This is represented in Fig.~\rf{f13}.
\bfg
\vspace*{0.25 cm}
\bc
\epsfig{file=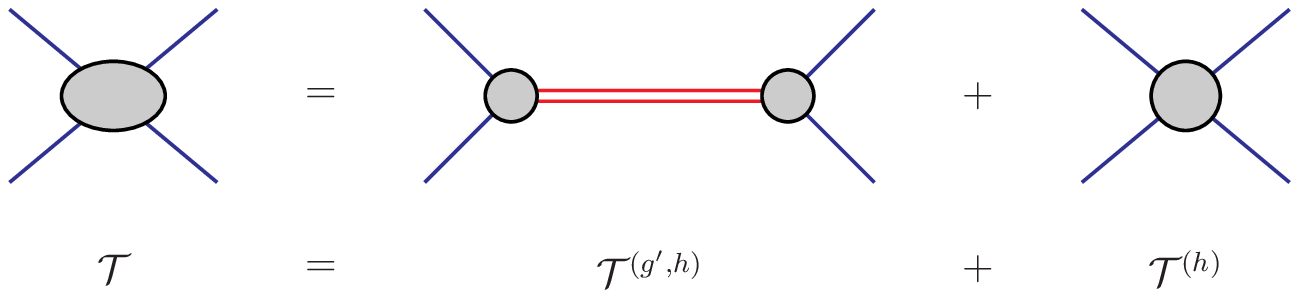,scale=0.8}
\vspace*{0.25 cm}
\caption{The meson-meson scattering amplitude, $\mathcal{T}$, due to
the contributions of the renormalized compact tetraquark pole, with
its renormalized vertices with two mesons, $\mathcal{T}^{(g',h)}$,
and the chain of contact interactions of Fig.~\rf{f2},
$\mathcal{T}^{(h)}$.}
\lb{f13}
\ec
\efg
\par
One obtains
\bea \lb{6e9}
\mathcal{T}&\equiv&\mathcal{T}^{(g',h)}+\mathcal{T}^{(h)}\nonumber \\
&=&-\frac{(m_1^{}+m_2^{})^2g^{\prime \mathrm{r}2}}
{(1-ih^{\mathrm{r}}J^{\mathrm{r}}(s))^2}\
\frac{1}{\Big[s-m_{tc1}^2+i(m_1^{}+m_2^{})^2
g^{\prime \mathrm{r}2}J^{\mathrm{r}}(s)/
(1-ih^{\mathrm{r}}J^{\mathrm{r}}(s))\Big]}\nonumber \\
& &\ \ \ \ \ \ 
+\frac{h^{\mathrm{r}}}{(1-ih^{\mathrm{r}}J^{\mathrm{r}}(s))}
\nonumber \\ 
&=&\frac{h^{\mathrm{r}}(s-m_{tc1}^2)
-(m_1^{}+m_2^{})^2g^{\prime \mathrm{r}2}}
{\Big[(s-m_{tc1}^2)(1-ih^{\mathrm{r}}J^{\mathrm{r}}(s))
+i(m_1^{}+m_2^{})^2g^{\prime \mathrm{r}2}J^{\mathrm{r}}(s)\Big]}
\eea
\par
The above expression could also have been obtained by starting
from the integral equation $i\mathcal{T}=K+iKJ\mathcal{T}$, with
the kernel $K$ given by
\be \lb{6e10}
K=-i\frac{(m_1^{}+m_2^{})^2g^{\prime 2}}{(s-m_{tc0}^2)}+ih.
\ee
After the renormalizations of the coupling constants and of the mass
$m_{tc0}^{}$ are done, one finds (\rf{6e9}).
\par

\subsection{Bound states} \lb{s62} 

The singularities of the scattering amplitude (\rf{6e9}) are the same
as those of the tetraquark propagator (\rf{6e8}). The separate
molecular-type singularity, present in $\mathcal{T}^{(h)}$, has been
cancelled by a similar singularity resulting from the radiative
corrections in $\mathcal{T}^{(g',h)}$.
\par
We first focus on the bound state problem. The tetraquark mass,
$m_t^{}$, is given by the equation
\be \lb{6e11}
(s_t^{}-m_{tc1}^2)(1-ih^{\mathrm{r}}J^{\mathrm{r}}(s_t^{}))
+i(m_1^{}+m_2^{})^2g^{\prime \mathrm{r}2}J^{\mathrm{r}}(s_t^{})=0,
\ee
where $s_t^{}=m_t^2$. Sticking to the nonrelativistic limit and
using definitions similar to (\rf{3e10}) and (\rf{3e10a}), and  
\bea \lb{6e12}
& &E_t^{}=m_t^{}-(m_1^{}+m_2^{}),\ \ \ \ \ \ \
e_t^{}\equiv\frac{E_t^{}}{2m_r^{}},\nonumber \\
& &E_{tc1}^{}=m_{tc1}^{}-(m_1^{}+m_2^{}),\ \ \ \ \ \ \
e_{tc1}^{}\equiv\frac{E_{tc1}^{}}{2m_r^{}},
\eea
Eq.~(\rf{6e11}) reduces to (omitting henceforth the
renormalization label r from the coupling constants)
\be \lb{6e13}
(-e_t^{}+e_{tc1}^{})(1+\frac{\alpha h}{16\pi}\sqrt{-e_t^{}})
+\frac{g^{\prime 2}}{16\pi}\sqrt{-e_t^{}}=0,
\ee
where $\alpha$ is defined in (\rf{3e11})\footnote{The presence of
the bare binding energy of the compact tetraquark, $-e_{tc1}^{}$,
introduces a new energy scale in the equations. Scaling $e_t^{}$
as $e_t^{}\rightarrow -e_{tc1}^{}e_t$ and the coupling constants
as $g^{\prime 2}\rightarrow \sqrt{-e_{tc1}^{}}g^{\prime 2}$,
$h\rightarrow h/\sqrt{-e_{tc1}^{}}$, one can get rid of 
${-e_{tc1}^{}}$ from the equation. We shall, nevertheless,
maintain the primary definitions, with the explicit presence of
${-e_{tc1}^{}}$, in order to remain closer to the physical
meaning of the different quantities.}.
We notice that in case $g'=0$ (absence of tetraquark-two-meson
coupling), the equation splits into two independent equations
yielding the mo\-le\-cular-type solution (\rf{3e11}) (for $h<0$),
on the one hand, and the bare compact tetraquark mass (\rf{6e7}),
on the other.
In case $h=0$ (absence of molecular-type forces), the equation
reduces to that of the compact tetraquark case (\rf{5e7}).
Equation (\rf{6e13}) cannot be solved analytically, but accurate
analytic approximate solutions can be found for it.
One has to distinguish two cases, according to the sign of $h$.
\par
We first consider the case $h>0$. According to the empirical
relationship (\rf{3e12}) and Fig. \rf{f3}, this case corresponds
to subcritical values of three-meson coupling constants, for which
no genuine molecular-type bound states can exist. Even though $h$
may take large values, approaching $+\infty$, one may characterize
this domain as globally representing a weak-coupling regime. One
therefore expects that the tetraquark bound state originates entirely
from a compact configuration, the molecular forces mainly introducing
deformations.
\par
A leading approximate solution to (\rf{6e13}) is
\be \lb{6e14}
\sqrt{-e_{t0}^{}}=\frac{1}{2}\Big[-b'+\sqrt{b^{\prime 2}
-4e_{tc1}^{}}\Big],
\ \ \ \ \ \ \ b'=\frac{g^{\prime 2}}{16\pi}
\frac{1}{(1+\frac{\alpha h}{16\pi}\sqrt{-e_{tc1}^{}})},
\ee
which generalizes (\rf{5e8}). 
This expression, together with its next-to-leading term, which is
presented in the Appendix \rf{a1} [Eq.~(\rf{ae1})], reproduces the
behavior of the exact solution with an error of less than a few
percent, the error slightly increasing with $h$.
The behavior of $\sqrt{-e_{t}^{}}$ with respect to variations of
$g^{\prime 2}/(16\pi)$, for fixed $h$, is similar to that of Fig.
\rf{f6}. With increasing $g^{\prime 2}/(16\pi)$, $\sqrt{-e_{t}^{}}$
approaches zero, starting from $\sqrt{-e_{tc1}^{}}$. The effect of the
presence of $h$ is simply the weakening of the slope of the decrease;
the molecular forces appear as opposed to the rapidity of the decrease.
Fig.~\rf{f14} displays the variation of $\sqrt{-e_{t}^{}}$ for a
few typical values of $\alpha h/(16\pi)$.
\bfg
\vspace*{0.25 cm}
\bc
\epsfig{file=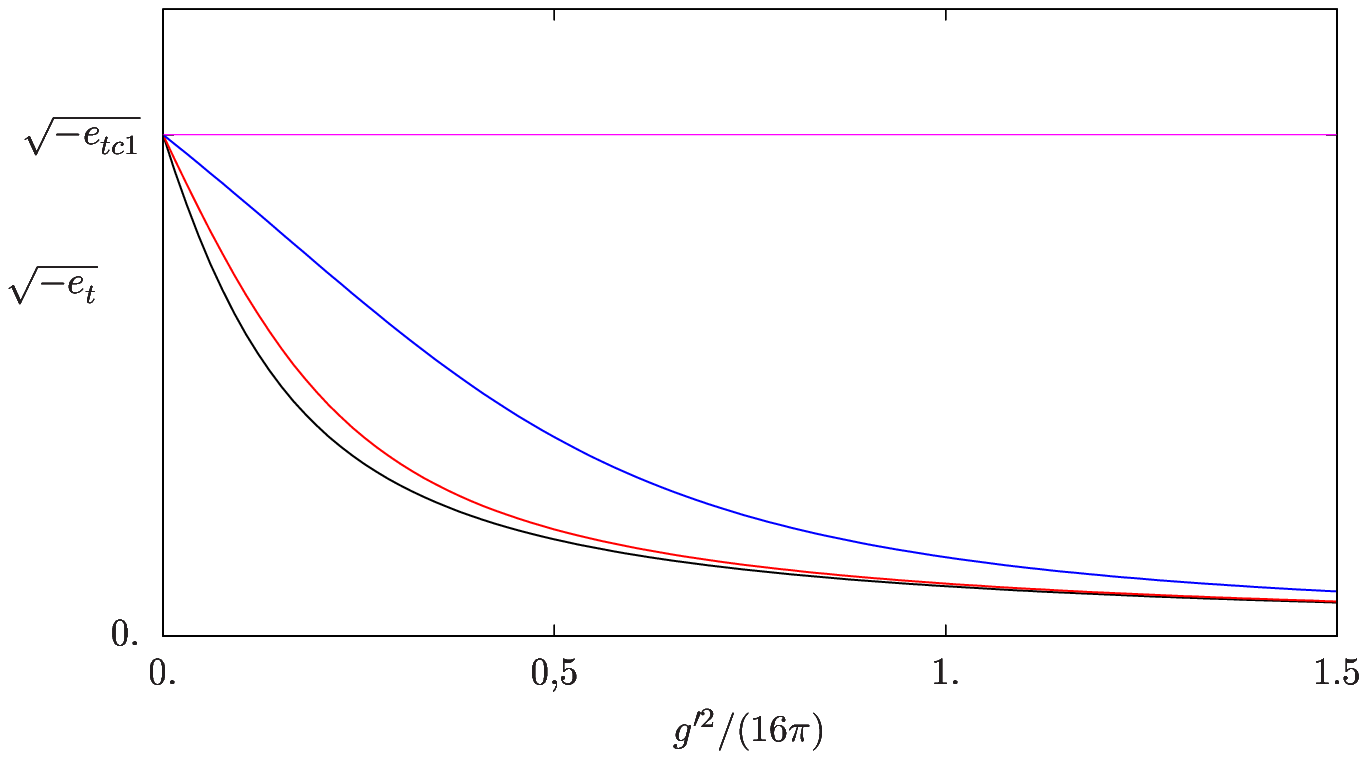,scale=0.8}
\vspace*{0.25 cm}
\caption{Variation of the square-root of the binding energy as a
function of $g^{\prime 2}/(16\pi)$, for four different positive
values of $\alpha h/(16\pi)$: $0$ (black curve), $5$ (red), $30$
(blue), $+\infty$ (magenta, the horizontal line at
$\sqrt{-e_{tc1}^{}}$); $-e_{tc1}^{}$ has been fixed at 0.01.}
\lb{f14}
\ec
\efg
\par
The second case to be considered is $h<0$. Here, we have two distinct
solutions, which we shall label with the indices 1 and 2,
corresponding to the generalizations of the two solutions existing in
the uncoupled case with $g'=0$. In the present domain of $h$, the most
interesting case corresponds to large values of $-h$, producing a
molecular-type state near the two-meson threshold. The case where $-h$
is small, actually corresponds, according to Fig.~\rf{f3}, to large
values of the three-meson coupling constant $g$, which lies outside
the domain of applicabilty of the nonrelativistic approximation.
Hence, we shall stick to a single large value of $-h$, fixed for
numerical applications at $-\alpha h/(16\pi)=10^{3/2}=31.62$, for
$-e_{tc1}=0.01$, and also shall consider its limiting value
$+\infty$. 
\par
The first solution describes the evolution of the molecular-type
solution (\rf{3e11}) under the influence of the coupling $g'$. 
A leading approximate expression of it is
\be \lb{6e15}
\sqrt{-e_{t0,(1)}^{}}=-\frac{e_{tc1}^{}}{\frac{g^{\prime 2}}{16\pi}
+e_{tc1}^{}\frac{\alpha h}{16\pi}}.
\ee
Its next-to-leading term is given in the Appendix \rf{a1}. The
analytic approximation (\rf{ae2}) reproduces the behavior of the
exact solution with an error of less than one per mil.
The behavior of $\sqrt{-e_{t,(1)}^{}}$ with respect to variations of
$g^{\prime 2}/(16\pi)$, for fixed $h$, is again similar to that of
Fig.~\rf{f6}; however, now, it starts, when $g^{\prime 2}/(16\pi)=0$,
from $\sqrt{-e_{tm}^{}}$ [Eq.~(\rf{3e11})], instead of
$\sqrt{-e_{tc1}^{}}$ (Fig.~\rf{f15}). In the limit 
$h\rightarrow -\infty$, $\sqrt{-e_{tm}^{}}$ tends to $0$ (the
two-meson threshold) and the whole curve coincides with the
horizontal $0$ line.
\par
The second solution describes the change of the compact-type
solution (\rf{5e8}) under the influence of the coupling $h$. 
A leading approximate expression of it is
\be \lb{6e16}
\sqrt{-e_{t0,(2)}^{}}=\sqrt{-e_{tc1}^{}-\frac{g^{\prime 2}/(16\pi)}
{\alpha h/(16\pi)}}.
\ee
(The condition $-\frac{\alpha h}{16\pi}\sqrt{-e_{t0,(2)}^{}}\gg 1$
should be fulfilled.)
Its next-to-leading term is given in the Appendix \rf{a1}. The
analytic approximation (\rf{ae3}) reproduces the behavior of the
exact solution with an error of less than one per cent.
The behavior of $\sqrt{-e_{t,(2)}^{}}$ with respect to variations of
$g^{\prime 2}/(16\pi)$, for fixed $h$, is represented by an increasing
function. The binding energy of the compact tetraquark thus increases
in the presence of the molecular-type state (Fig.~\rf{f15}). In the
limit $h\rightarrow -\infty$ the curve coincides with the horizontal
$\sqrt{-e_{tc1}^{}}$ line. Actually, the solution
$\sqrt{-e_{t,(2)}^{}}$ can be considered as the continuation of the
solution found in the case of positive values of $h$ to negative
values of $h$. As can be seen in Fig.~\rf{f14}, when $h$ increases
with positive values, the solution reaches, in the limit
$h\rightarrow +\infty$, the constant value $\sqrt{-e_{tc1}^{}}$.
Then, according to Fig.~\rf{f3}, $h$ passes to $-\infty$, which
also corresponds, for the solution $\sqrt{-e_{t,(2)}^{}}$, to the
same constant value $\sqrt{-e_{tc1}^{}}$ in Fig.~\rf{f15}. The
increasing of $h$ with negative values then produces the curves lying
above that horizontal line. 
\bfg
\vspace*{0.25 cm}
\bc
\epsfig{file=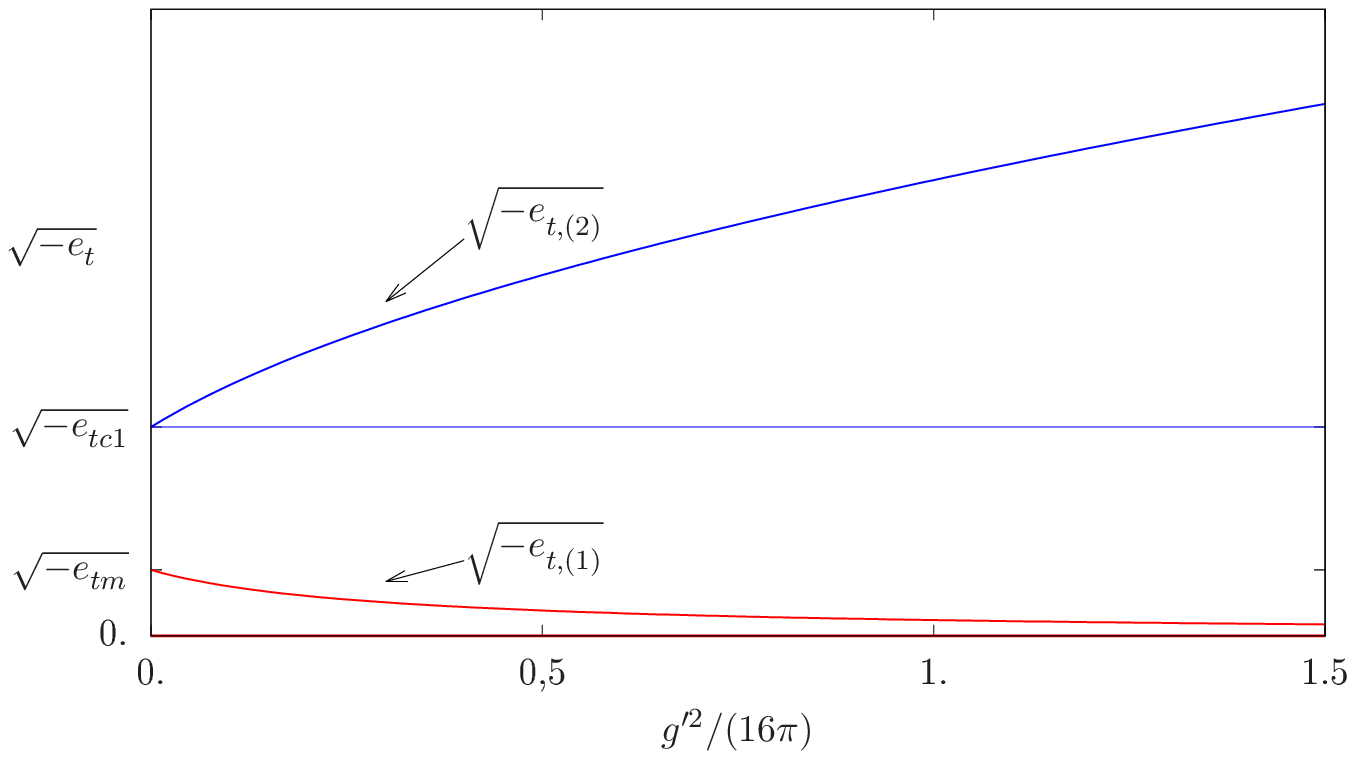,scale=0.8}
\vspace*{0.25 cm}
\caption{The two solutions of the bound state equation (\rf{6e13})
and their variation under changes of $g^{\prime 2}/(16\pi)$.
The value of $-\alpha h/(16\pi)$ has been fixed at $31.62$ and
that of $-e_{tc1}^{}$ at $0.01$. At the limiting value $h=-\infty$,
$\sqrt{-e_{t,(1)}^{}}$ coincides with the horizontal line $0$,
while $\sqrt{-e_{t,(2)}^{}}$ coincides with the horizontal line
$\sqrt{-e_{tc1}^{}}$.}
\lb{f15}
\ec
\efg
\par
Expanding the scattering amplitude (\rf{6e9}) around the bound state
pole (cf. (\rf{3e14})), one obtains the physical coupling constant
$g_{TM_1^{}M_2^{}}^{}$, expressed in two equivalent ways, using
(\rf{6e13}):
\be \lb{6e17}
g_{TM_1^{}M_2^{}}^2=\frac{32\pi\sqrt{-e_t^{}}}
{\Big[1+\frac{2g^{\prime 2}}{16\pi}\frac{(-e_t^{})^{3/2}}
{(e_t^{}-e_{tc1}^{})^2}\Big]}=      
\frac{32\pi\sqrt{-e_t^{}}}
{\Big[1+\frac{2\sqrt{-e_t^{}}}{g^{\prime 2}/(16\pi)}
\Big(1+\frac{\alpha h}{16\pi}\sqrt{-e_t^{}}\Big)^2\Big]}.
\ee
Comparison of these expressions with (\rf{5e2}) yields $Z$:
\be \lb{6e18}
Z=\frac{\frac{2g^{\prime 2}}{16\pi}(-e_t^{})^{3/2}}
{(e_t^{}-e_{tc1}^{})^2+\frac{2g^{\prime 2}}{16\pi}(-e_t^{})^{3/2}}
=\frac{\Big(1+\frac{\alpha h}{16\pi}\sqrt{-e_t^{}}\Big)^2
\sqrt{-e_t^{}}}{\Big(1+\frac{\alpha h}{16\pi}\sqrt{-e_t^{}}\Big)^2
\sqrt{-e_t^{}}+\frac{1}{2}\frac{g^{\prime 2}}{16\pi}},
\ee
which is manifestly a positive quantity bounded by 1.
\par
The value of $Z$ and its behavior under variations of the coupling
constants depend on the specific bound states that we have met above. 
When $h>0$, we have one bound state with the square-root of the
binding energy having the approximate expressions (\rf{6e13}) and
(\rf{ae1}). For this
case, the qualitative features of $Z$ are more transparent in the
second expression of (\rf{6e18}). When $g'\rightarrow 0$, for fixed
$h$, $\sqrt{-e_t^{}}\rightarrow \sqrt{-e_{tc1}^{}}$ (cf.
Fig.~\rf{f14}); then $Z\rightarrow 1$. When $g'$ increases,
$\sqrt{-e_t^{}}$ decreases rapidly and tends to zero. The behavior
of $Z$ remains very similar to that of Fig.~\rf{f8}, with the
difference that the presence of $h$ slows down the decrease of $Z$.
We display, in Table \rf{t1}, the values of $Z$ for several values
of $\alpha h/(16\pi)$, for
$g^{\prime 2}/(16\pi)=0.5$ and $-e_{tc1}^{}=0.01$.
\begin{table}[ht]
\bc
\caption{Values of $Z$, for several positive values of
$\alpha h/(16\pi)$ (Fig. \rf{f14}), for
$g^{\prime 2}/(16\pi)=0.5$ and $-e_{tc1}^{}=0.01$.}
\lb{t1}
\begin{tabular}{cccccc}
\hline
$\alpha h/(16\pi)$ & 0 & 1. & 5. & 30. & $\infty$ \\
\hline
$Z$	& 0.072 & 0.075 & 0.094 & 0.388 & 1. \\
\hline
\end{tabular}
\ec
\end{table}
\par
When $h<0$, we have two bound states with the square-root of the
binding energies having the approximate expressions (\rf{6e15}) and
(\rf{ae2}), on the one hand, and (\rf{6e16}) and (\rf{ae3}), on the
other. For the solution 1, $\sqrt{-e_t^{}}$ remains much smaller
than $\sqrt{-e_{tc1}^{}}$ (cf. Fig.~\rf{f15}); in that case, the
first expression of $Z$ in (\rf{6e18}) is more adequate for the
analysis. When $g'\rightarrow 0$, $Z\rightarrow 0$. However, the
evolution of $Z$ under variations of $g'$ is no longer monotonic.
For fixed $h$, $Z$ increases, starting from zero at $g'=0$, reaches
a maximum value nearly at
$g^{\prime 2}/(16\pi)=e_{tc1}^{}\alpha h/(32\pi)$
($\simeq 0.16$ for $\alpha h/(16\pi)=-31.62$ and $-e_{tc1}^{}=0.01$),
then decreases down to 0. The bound state remains, therefore, very
close to a molecular configuration in all the interval of variation
of $g'$. The main influence of the latter is reflected in the
continuous decrease of the binding energy, accentuating the
shallowness of the state. We display, in Table \rf{t2}, for
$\alpha h/(16\pi)$ fixed at $-31.62$, with $-e_{tc1}^{}=0.01$,  
the values of $Z$ for several values of $g^{\prime 2}/(16\pi)$.
\begin{table}[ht] 
\bc
\caption{Values of $Z$, corresponding to solution 1 of Fig.
\rf{f15}, for several values of $g^{\prime 2}/(16\pi)$,
for $\alpha h/(16\pi)=-31.62$, with $-e_{tc1}^{}=0.01$.}
\lb{t2}
\begin{tabular}{ccccccccc}
\hline
$g^{\prime 2}/(16\pi)$ & 0 & 0.1 & 0.16 & 0.25 & 0.5 & 1. & 1.5 &
$\infty$ \\
\hline
$Z$	& 0 & 0.029 & 0.030 & 0.027 & 0.018 & 0.008 & 0.005 & 0.  \\
\hline
\end{tabular}
\ec
\end{table}
\par
For the solution 2, the second expression of $Z$ in (\rf{6e18})
is more adequate for the analysis. When $g'\rightarrow 0$,
$Z\rightarrow 1$ (the factor $(1+\alpha h/(16\pi)\sqrt{-e_t^{}})$
does not vanish).  
When $g'$ increases, $Z$ decreases, reaches a minimum value nearly
at $g^{\prime 2}/(16\pi)=2e_{tc1}^{}\alpha h/(16\pi)$  
($\simeq 0.6$ for $\alpha h/(16\pi)=-31.62$ and $-e_{tc1}^{}=0.01$),
then increases up to 1. The bound state remains very close to the
compact configuration in all the interval of variation of $g'$.
The main influence of the latter is reflected in the continuous
increase of the binding energy (Fig.~\rf{f15}). 
We display, in Table \rf{t3}, for $\alpha h/(16\pi)$ fixed at
$-31.62$, with $-e_{tc1}^{}=0.01$, the values of $Z$ for several
values of $g^{\prime 2}/(16\pi)$.
\begin{table}[ht] 
\bc
\caption{Values of $Z$, corresponding to solution 2 of Fig.
\rf{f15}, for several values of $g^{\prime 2}/(16\pi)$, for
$\alpha h/(16\pi)=-31.62$, with $-e_{tc1}^{}=0.01$.}
\lb{t3}
\begin{tabular}{cccccccc}
\hline
$g^{\prime 2}/(16\pi)$ & 0 & 0.1 & 0.25 & 0.5 & 1. & 1.5 & $\infty$ \\
\hline
$Z$	& 1. & 0.948 & 0.933 & 0.930 & 0.937 & 0.943 & 1. \\
\hline
\end{tabular}
\ec
\end{table}
\par
To determine the scattering length and the effective range, one
expresses the scattering amplitude $\mathcal{T}$ in the scattering
region $E>0$:
\be \lb{6e19}
\mathcal{T}=8\pi(m_1^{}+m_2^{})\,\Big[\,\frac{32\pi(e-e_{tc1}^{})
m_r^{}}{(e-e_{tc1})\alpha h-g^{\prime 2}}+\frac{(1-d)k^2}{\pi m_r^{}}
-ik\,\Big]^{-1},  
\ee
from which one deduces, with the aid of (\rf{4e2}) and (\rf{4e3}),
\be \lb{6e20}
a=\frac{1}{2m_r^{}}\Big(-\frac{\alpha h}{16\pi}-\frac{1}{e_{tc1}^{}}
\frac{g^{\prime 2}}{16\pi}\Big),\ \ \ \ \ \
r_e^{}=-\frac{1}{m_r^{}}\frac{g^{\prime 2}/(16\pi)}
{\Big(\frac{\alpha h}{16\pi}e_{tc1}^{}+\frac{g^{\prime 2}}{16\pi}
\Big)^2}+\frac{1}{m_r^{}}\frac{2}{\pi}(1-d),
\ee
where $d$ is defined in (\rf{4e5}) and $\alpha$ in (\rf{3e11}). 
The relationship of $a$ and $r_e^{}$ with $Z$ [Eq. (\rf{6e18})]
is not, however, as simple as in the previous cases
[Eqs.~(\rf{5e1}) and (\rf{5e31})]. The reason for this comes from
the fact that $\mathcal{T}$ contains now a zero in the vicinity of
the bound state pole and does not satisfy Weinberg's representation
(\rf{5e2b}) \cite{Castillejo:1955ed,Oller:1998zr}.
For $h>0$, the zero occurs in the domain $e>e_{tc1}^{}$, 
while for $h<0$, it occurs in the domain $e<e_{tc1}^{}$.
Nevertheless, expressions (\rf{6e20}) are simple enough
in terms of the elementary parameters of the theory, and together
with the knowledge of the binding energy, if they are measured
experimentally or on the lattice, allow the calculation of $Z$.
In particular, the presence of the coupling $g'$ between the
tetraquark and the two mesons provides $r_e^{}$ in general with
a negative value, whatever the sign of $h$ is. We notice that
for $h<0$, $a$ is positive, as it receives contributions from two
independent bound states. For $h>0$, $a$ is positive for
$\alpha h<-g^{\prime 2}/e_{tc1}^{}$ and negative for
$\alpha h>-g^{\prime 2}/e_{tc1}^{}$. For
$\alpha h=-g^{\prime 2}/e_{tc1}^{}$, $a$ vanishes and simultaneously
$r_e^{}$ tends to $-\infty$. For $g^{\prime 2}/(16\pi)=0.5$ and
$-e_{tc1}^{}=0.01$, the latter critical value of $h$ occurs at
$\alpha h/(16\pi)=50$. (Cf. (\rf{3e12}) for its conversion into
a three-meson coupling constant $\overline g$; the latter lies
slightly below the critical value $\overline g_{\mathrm{cr}}^{}$
displayed in (\rf{2e7}).)
\par

\subsection{Resonances} \lb{s63} 

Resonances may occur when the renormalized (real) compact tetraquark
mass, given by (\rf{6e7}), lies above the two-meson threshold. In that
case, the nonrelativistic energy $E_{tc1}^{}$ of (\rf{6e12}) is positive.
The poles of the scattering amplitude (\rf{6e9}) may have complex
values. Designating by $E_{TR}^{}$ the complex energy of the pole,
the equivalent of Eq.~(\rf{6e13}) is now
\be \lb{6e21}
(e_{TR}^{}-e_{tc1}^{})(1-i\frac{\alpha h}{16\pi}\sqrt{e_{TR}^{}})
+i\frac{g^{\prime 2}}{16\pi}\sqrt{e_{TR}^{}}=0.
\ee
When $h\neq 0$, this equation has to be solved numerically.
Introducing the definitions
\be \lb{6e22}
\sqrt{e_{TR}^{}}=u+i\,v,\ \ \ \ 
e_{TR}^{}=u^2-v^2+i\,2uv\equiv e_{TRr}^{}+i\,e_{TRi}^{},
\ \ \ \ \ \ u,v\ \mathrm{real},
\ee
where $e_{TRr}^{}$ and $e_{TRi}^{}$ are the real and imaginary parts
of $e_{TR}^{}$, respectively, one can study the behavior of 
$e_{TR}^{}$ under variations of $g^{\prime 2}$ and $h$. The
physical conditions to be imposed on the solutions are $v<0$,
$e_{TRi}^{}<0$ and  $e_{TRr}^{}>0$ (the resonance lies above the
two-meson threshold). In the case $h=0$, the upper bound
(\rf{5e20}) had been found for $g^{\prime 2}/(16\pi)$.
\par
When $h\neq 0$, one has to distinguish two main domains of $h$
(cf. Fig.~\rf{f3}): (i) $h>0$ and small, corresponding to the
weak-coupling regime of the meson-meson interaction; (ii) $|h|$
large, corresponding to the strong-coupling regime, with the
possibility of existence of a genuine molecular-type bound state
(for $h<0$). In the first case, the qualitative behavior of the
solutions is naturally close to that of the case with $h=0$,
studied in Sec.~\rf{s5}, the only changes being small
quantitative ones. Thus, for $h/(16\pi)\leq 0.2/\sqrt{-e_{tc1}^{}}$,
a conservative upper bound for $g^{\prime 2}$
is $g^{\prime 2}/(16\pi)\leq\sqrt{-e_{tc1}^{}}$. When $g^{\prime 2}$
increases from 0 to its upper bound, the resonance approaches the
two-meson threshold with its imaginary part remaining finite, but
relatively decreasing. In the second case, for the same variation
of $g^{\prime 2}$, the resonance approches the threshold for $h<0$
and moves away from it for $h>0$. There are also regions of $h$
for which large values of $g^{\prime 2}$ are possible, however
they are not obtained by continuous variations of all interactions.
\par
For completeness, we display here the equation satisfied by the
imaginary part of $\sqrt{e_{TR}^{}}$:
\be \lb{6e23}
v=-\frac{1}{2}\,\frac{g^{\prime 2}}{16\pi}\,
\frac{1}{\Big(1+2v\frac{\alpha h}{16\pi}+(\frac{\alpha h}
{16\pi})^2|e_{TR}^{}|\Big)},
\ee
$|e_{TR}^{}|$ being the modulus of $e_{TR}^{}$, which generally
favors negative values of $v$ (when $h<0$ or when $h>0$ but small).
\par
Expanding $\mathcal{T}$ [Eq.~(\rf{6e9})] around $E_{TR}^{}$,
as in (\rf{5e22}), one obtains the expression of the 
tetraquark-two-meson coupling constant squared:
\be \lb{6e24}
g_{TM_1^{}M_2^{}}^2=32\pi\sqrt{e_{TR}^{}}\,
\frac{g^{\prime 2}/(16\pi)}{\Big[2\sqrt{e_{TR}^{}}
\Big(1-i\frac{\alpha h}{16\pi}\sqrt{e_{TR}^{}}\Big)^2
+i\frac{g^{\prime 2}}{16\pi}\Big]}.
\ee
Generally, the multiplicative factor accompanying
$32\pi\sqrt{e_{TR}^{}}$ is identified with the compositeness
coefficient $(1-Z)$ [Eqs.~(\rf{5e2}), (\rf{5e24}), (\rf{6e18})].
Here, however, the corresponding coefficient is complex when
$h\neq 0$ and, therefore, a probabilistic interpretation of it is
no longer possible (cf. also \cite{Sekihara:2014kya,Guo:2015daa}).
A natural extension of the usual definition would correspond to
taking the modulus of the corresponding expression as being
equivalent to $(1-Z)$. While such an extension would ensure the
reality condition of the probability candidate, it does not yet
guarantee its boundedness by 1.
Indeed, one may check that when the coupling
constant $g^{\prime 2}$ exceeds its upper bound, mentioned earlier in
this subsection, one finds, for small positive values of $h$, that
$(1-Z)$ exceeds 1. Such a situation has also been found in
Sec.~\rf{s53} and has been interpreted as a sign of the instability
of the initial system, leading to the disappearance of the compact
tetraquark from the spectrum. On the other hand, large values of
$g^{\prime 2}$ generally send back the resonance to the bound state
domain. We therefore adopt the modulus prescription, with the
restriction that $g^{\prime 2}$ respects its upper bound:
\be \lb{6e25}
(1-Z)=\Big|\frac{g^{\prime 2}/(16\pi)}{\Big[2\sqrt{e_{TR}^{}}
\Big(1-i\frac{\alpha h}{16\pi}\sqrt{e_{TR}^{}}\Big)^2
+i\frac{g^{\prime 2}}{16\pi}\Big]}\Big|.
\ee
The general properties of $Z$ are then similar to those met in the
case $h=0$. When the resonance approaches the two-meson threshold,
$Z$ tends to 0, while when the resonance stays in the vicinity of its
primary position, this mainly corresponding to small values of
$g^{\prime 2}$, $Z$ remains close to 1. On the other hand, large and
negative values of $h$ have the tendency to push the resonance towards
the two-meson threshold, while large and positive values of $h$ repel
the resonance from the threshold.
\par
It is worthwile recalling that in the case $h<0$, the spectrum also
contains a molecular-type bound state, whose binding energy has been
approximately estimated, in the two-bound-state case, by means of
Eq.~(\rf{6e15}), where $e_{tc1}<0$. The same formula could also
be used for the evaluation of the binding energy in the resonance
case, where now $e_{tc1}>0$:
\be \lb{6e26}
\sqrt{-e_{t0,(1)}^{}}=\frac{1}{\Big(-\frac{\alpha h}{16\pi}
-\frac{g^{\prime 2}}{16\pi e_{tc1}^{}}\Big)}.
\ee
We find that the bound state exists only when $-{\alpha h}/(16\pi)$
is greater than ${g^{\prime 2}}/(16\pi e_{tc1}^{})$; the condition of
the vicinity of the bound state to the two-meson threshold actually
requires much larger values. For instance, with $e_{tc1}^{}=0.01$ and
${g^{\prime 2}}/(16\pi)=0.1$, one would need $-{\alpha h}/{16\pi}>10$.
Values of $-{\alpha h}/{16\pi}$ of the order of 30, which were
frequently considered throughout the present work, would then produce
a binding energy $-e_{t0,(1)}$ of the order of 0.0025.
\par
The expressions of the scattering length and the effective range are
the same as in Eqs.~(\rf{6e20}), with the only difference that
$e_{tc1}$ is now positive. The contribution of the term proportional
to $g^{\prime 2}$ in $a$ is now negative and in the case $h<0$
the same competition as in the case of the bound-state binding
energy (\rf{6e26}) arises between the contributions of $h$ and
$g^{\prime 2}$. In case
$-\alpha h/(16\pi)-g^{\prime 2}/(16\pi e_{tc1}^{})$ is negative,
the bound state disappears and the scattering length becomes negative.
While the effective range is insensitive to the sign of that
combination, it nevertheless has a singularity when the latter
vanishes.
\par
In summary, the existence of resonances is much sensitive to the
strength of the primary (bare) tetraquark-two-meson coupling constant,
which should remain sufficiently weak. Large values of it generally
resend the resonance to the bound state region. There is also an
interval of the coupling constant for which the tetraquark may
disappear from the spectrum. This is in contrast to the bound state
case, where all values of the coupling constant are acceptable, with
different consequences according to the values of the four-meson
coupling constant.
\par

\section{Large-{\boldmath $N_{\lowercase{c}}^{}$} analysis}
\lb{s7}

At large $N_c^{}$, ordinary mesons are stable noninteracting
particles
\cite{'tHooft:1973jz,'tHooft:1974hx,Callan:1975ps,Witten:1979kh,
Witten:1979pi,Coleman:1985rnk} and can be considered as
compact objects. Their couplings to other mesons is of subleading
order in $N_c^{}$. Therefore, the latter can act on them only as
perturbations, not affecting their compact structure.
\par
This is not the case of tetraquarks, because of the existence of
internal mesonic clusters. The compact structure of primarily
existing tetraquarks is not protected by the large-$N_c^{}$ limit.
This is due to the fact that the interaction forces acting for the
formation of mesonic clusters are $(N_c^{}-1)$ times larger than
the forces forming diquark compact objects
\cite{Esposito:2016noz,Lucha:2021mwx} (assuming that the confining
forces have the same color structure as one-gluon exchange terms). 
This means that the primary coupling constant  $g^{\prime}$
that connects the compact tetraquark to the two mesons should
behave at large $N_c^{}$ like $N_c^{1/2}$, assuming that the
compact tetraquark state and energy are of order $N_c^{0}$.
For the squared quantity, one has
\be \lb{7e1}
g^{\prime 2}=O(N_c^{}).
\ee
\par
Concerning the four-meson contact-type coupling constant $h$, we
have emphasized in Sec. \rf{s3} that it is not an elementary
coupling constant and should rather be related to the three-meson
coupling constants of the higher-energy theory. The latter coupling
constants generically behave, at large $N_c^{}$, like $N_c^{-1/2}$
\cite{Witten:1979kh,Coleman:1985rnk,Callan:1975ps,Lucha:2021mwx}
and vanish in that limit. Compared to the critical coupling constant,
introduced in (\rf{2e7}), they lie, in the subcritical domain.
Meson-meson interactions cannot, therefore, produce on their own
bound states in the large-$N_c^{}$ limit.
Going back to the empirical formula (\rf{3e12}), we deduce
that $h$ is positive and lies in its perturbative domain, behaving
like $\overline g^2$:
\be \lb{7e2}
h=O(N_c^{-1}),\ \ \ \ \ h>0.
\ee
The meson-meson interaction has, therefore, only a subleading effect
with respect to the direct compact-tetraquark--two-meson interaction.
From Eqs.~(\rf{6e14}) and (\rf{5e10}) one deduces that the
tetraquark binding energy vanishes like $N_c^{-2}$:
\be \lb{7e3}
-e_{tc}^{}=O(N_c^{-2}).
\ee
From (\rf{6e18}) and (\rf{5e14}), one also finds the behavior of
the elementariness coefficient:
\be \lb{7e4}
Z=O(N_c^{-2}).
\ee
At large $N_c^{}$, the compact tetraquark is thus transformed into
a shallow, molecular-type, bound state.
\par
From Eqs.~(\rf{6e17}) and (\rf{5e12}), one deduces the behavior
of the physical coupling constant $g_{TM_1^{}M_2^{}}^{}$:
\be \lb{7e5}
g_{TM_1^{}M_2^{}}^{2}=O(N_c^{-1}).
\ee
This result is in accordance with other general estimates
done in the large-$N_c^{}$ limit, which predict the vanishing of
the coupling constant in that limit
\cite{Weinberg:2013cfa,Knecht:2013yqa,Cohen:2014tga,Maiani:2016hxw,
Maiani:2018pef,Lucha:2017mof,Lucha:2017gqq,Lucha:2021mwx}.
The behavior (\rf{7e5}) should only be considered as a generic one.
The power of the decrease may slightly change according to the
detailed flavor content of the tetraquark state or other more
refined analyses.
\par
Concerning the resonance states, we had found that they show up
only in the weak-coupling regime. However, the large-$N_c^{}$
limit imposes the strong-coupling regime. Therefore, in that limit,
resonances should disappear from the spectrum, at least from the
neighberhood of the two-meson threshold. 
\par
In conclusion, in the large-$N_c^{}$ landscape, tetraquarks may
survive only in the form of shallow, molecular-type, bound states,
which are relics of primarily created compact states.
\par

\section{Conclusion} \lb{s8}

The compact tetraquark scheme, considered in its simplest version,
where di\-quarks and anti\-diquarks are separately gathered in very
small volumes, can be considered as a starting point for the
analysis of the tetraquark properties. In this situation, because
of the dominance of the attractive confining forces, one usually
finds confined bound states, in parallel to the case of ordinary
hadrons. However, the very existence of underlying meson-clustering
interactions in the general system forces the initial compact state
to evolve towards a more complicated structure, where now
molecular-type configurations are also present. In an effective
field theory approach, where the compact tetraquark is represented
as an elementary particle, this evolution was studied, within
a scalar interaction framework, by means
of the primary compact-tetraquark--two-meson coupling constant.
Another quantity, more related to physical observables, is the
elementariness coefficient $Z$, which varies between 1, corresponding
to the pure elementary case, and 0, corresponding to the completely
composite case. Stronger is the primary coupling constant, smaller is
the value of $Z$. In the strong-coupling limit, the system tends
to a dominant molecular configuration, characterized by a shallow
structure. In the case of resonances, only the weak-coupling regime
provides a stable framework for their existence. For higher values
of the coupling constant, either the resonance disappears from the
spectrum, or is resent to the bound state domain.
\par
The consideration of the large-$N_c^{}$ limit of QCD provides an
additional support to the dominance of the strong-coupling regime
of the coupling constant, with all its consequences.
\par
Many of the tetraquark candidates, whose elementariness has been
evaluated in the literature from experimental data, have led to
values of $Z$ which are neither 1, nor 0. This is a clear indication
of the mixture of configurations that results from the evolution
described above. Nonzero values of $Z$, even small ones, are
indicative of the existence of a primary compact state. Shallowness
of bound states, with small values of $Z$, may receive a natural
explanation as resulting from the strong-coupling limit of the
interaction compact-tetraquark--meson-clusters, also supported by
the large-$N_c^{}$ limit of QCD.
\par
The analysis undertaken in the present work was based on the
simplest qualitative approach, considering scalar interactions,
ignoring spin degrees of freedom and details of quark flavors, and
using nonrelativistic limit and single-channel formalism for the
meson clusters. A more general and refined quantitative analysis,
concerning definite candidates, should include the missing
ingredients.
\par

\begin{acknowledgments}
The author thanks Jaume Carbonell, Marc Knecht,
Wolfgang Lucha, Dmitri Melikhov, Bachir Moussallam, and Ubirajara
van Kolck for useful discussions.  
This research received financial support from the EU
research and innovation programme Horizon 2020, under Grant
agree\-ment No.~824093, and from the joint CNRS/RFBR Grant
No.~PRC Russia/19-52-15022.
\end{acknowledgments}

\appendix*
\section{} \lb{a1}
%\subsection{}
We present in this appendix the approximate analytic expressions of
the solutions of Eq.~(\rf{6e13}) for the two cases of $h$.
\par
For the case $h>0$, the approximate solution, including also the
next-to-leading term to (\rf{6e14}), is
\be \lb{ae1}
\sqrt{-e_{t}^{}}\simeq\sqrt{-e_{t0}^{}}\Big[1-
\frac{b'\frac{\alpha h}{16\pi}(\sqrt{-e_{tc1}^{}}-\sqrt{-e_{t0}^{}})}
{\Big(\frac{g^{\prime 2}}{16\pi}-\frac{2\alpha h}{16\pi}e_{tc1}^{}
+(2-3b'\frac{\alpha h}{16\pi})\sqrt{-e_{t0}^{}}\Big)}\Big].         
\ee
\par
For the case $h<0$, the first solution, in its approximate form,
including also the next-to-leading term to (\rf{6e15}), is
\be \lb{ae2}
\sqrt{-e_{t,(1)}^{}}\simeq\sqrt{-e_{t0,(1)}^{}}\Big[1-
\frac{\sqrt{-e_{t0,(1)}^{}}(1+\frac{\alpha h}{16\pi}
\sqrt{-e_{t0,(1)}^{}})}{\Big(\frac{g^{\prime 2}}{16\pi}+
\frac{\alpha h}{16\pi}e_{tc1}^{}+\sqrt{-e_{t0,(1)}^{}}
(2+3\frac{\alpha h}{16\pi}\sqrt{-e_{t0,(1)}^{}})\Big)}\Big].
\ee
\par
The second solution, in its approximate form,
including also the next-to-leading term to (\rf{6e16}), is
\be \lb{ae3}
\sqrt{-e_{t,(2)}^{}}\simeq\sqrt{-e_{t0,(2)}^{}}
-\frac{g^{\prime 2}/(16\pi)}
{2\alpha h/(16\pi)}\ \frac{1}{\Big(\frac{g^{\prime 2}}{16\pi}+
\frac{\alpha h}{16\pi}e_{tc1}^{}-\sqrt{-e_{t0,(2)}^{}}\Big)}.
\ee
\par     

\bibliography{qrmtq-v2-arx.bib}

\end{document}